\documentclass[notitlepage,aps,prd,twocolumn,reprint,superscriptaddress,showpacs]{revtex4-1}

\usepackage{natbib}
\usepackage[english]{babel}
\usepackage{graphicx}

\usepackage{amsmath, amssymb}
\usepackage{picinpar}
\usepackage{mathtools}
\usepackage[pdfborderstyle={/S/U/W 1}]{hyperref}
\usepackage{bbold}
\usepackage{color}
\usepackage{empheq}
\usepackage{color,soul}
\usepackage{siunitx}
\usepackage{tensor}
\usepackage{latexsym}
\usepackage{placeins}
\usepackage{csquotes}
\usepackage[acronym]{glossaries}

\usepackage{xfrac}
\usepackage{multirow}
\usepackage{titlesec}

\DeclareMathSizes{10}{10}{8}{6}
\DeclareSIUnit{\persqrthz}{\ensuremath{\text{\hertz}^{-1/2}}}
\renewcommand{\arraystretch}{1.5}

\titleformat{\section}[runin]{\normalfont\bfseries}{\thesection.}{0.5em}{}[ -]
\titlespacing{\section}{0pc}{5mm}{2mm}

\everymath{\displaystyle}

\begin{document}

%\title{LISA Pathfinder drag-free performances: \\Frequency domain analysis of the in-loop full dynamics of the test masses}
\title{LISA Pathfinder Platform Stability and Drag-free Performance}

\graphicspath{{./Images/}}

\newcommand{\indice}[1]{{\scriptscriptstyle #1}}
\newcommand{\exposant}[1]{{\scriptscriptstyle #1}}
\newcommand{\myvec}[2]{\vec{#1}_\indice{#2}}
\newcommand{\myexpr}[3]{#1_\indice{#2}^\exposant{#3}}
\DeclarePairedDelimiterX{\norm}[1]{\lVert}{\rVert}{#1}
\newcommand{\ddt}[2]{ \frac{d}{dt} \left( #1 \right)_#2}
\newcommand{\ddtddt}[2]{ \frac{d^2}{dt^2} \left( #1 \right)_#2}
\newcommand{\myhyperref}[1]{\hyperref[#1]{\ref{#1}}}
\newcommand{\identite}[1]{{\displaystyle \mathbb{1}_{\indice{#1}}}}
\newcommand{\myskew}[2][]{\@ifmtarg{#1}{\left[ #2 \right]^{\times}}{\left[ #2 \right]^{\times, #1}}}
\newcommand{\myat}[2][]{#1|_{#2}}
\newcommand{\timestentothe}[1]{\times 10^{#1}}
\renewcommand{\arraystretch}{1.5}

\makeatletter

% Search and replace example with OgreKit:
% Search: \\vec{(?<i>\w)}_\\indice{(?<o>\w/\w)}
% Replace: \\myvec{\g<i>}{\g<o>}

\makeatother

%\preprint{}


%\author{LISA Pathfinder Collaboration}
%\thanks{}
%\affiliation{}


\def\addressa{European Space Astronomy Centre, European Space Agency, Villanueva de la
Ca\~{n}ada, 28692 Madrid, Spain}
\def\addressb{Albert-Einstein-Institut, Max-Planck-Institut f\"ur Gravitationsphysik und Leibniz Universit\"at Hannover,
Callinstra{\ss}e 38, 30167 Hannover, Germany}
\def\addressc{APC, Univ Paris Diderot, CNRS/IN2P3, CEA/lrfu, Obs de Paris, Sorbonne Paris Cit\'e, France}
\def\addressd{High Energy Physics Group, Physics Department, Imperial College London, Blackett Laboratory, Prince Consort Road, London, SW7 2BW, UK }
\def\addresse{Dipartimento di Fisica, Universit\`a di Roma ``Tor Vergata'',  and INFN, sezione Roma Tor Vergata, I-00133 Roma, Italy}
\def\addressf{Department of Industrial Engineering, University of Trento, via Sommarive 9, 38123 Trento, 
and Trento Institute for Fundamental Physics and Application / INFN}
\def\addressh{European Space Technology Centre, European Space Agency, 
Keplerlaan 1, 2200 AG Noordwijk, The Netherlands}
\def\addressi{Dipartimento di Fisica, Universit\`a di Trento and Trento Institute for 
Fundamental Physics and Application / INFN, 38123 Povo, Trento, Italy}
\def\addressk{Istituto di Fotonica e Nanotecnologie, CNR-Fondazione Bruno Kessler, I-38123 Povo, Trento, Italy}
\def\addressj{The School of Physics and Astronomy, University of
Birmingham, Birmingham, UK}
\def\addressl{Institut f\"ur Geophysik, ETH Z\"urich, Sonneggstrasse 5, CH-8092, Z\"urich, Switzerland}
\def\addressm{The UK Astronomy Technology Centre, Royal Observatory, Edinburgh, Blackford Hill, Edinburgh, EH9 3HJ, UK}
\def\addressn{Institut de Ci\`encies de l'Espai (CSIC-IEEC), Campus UAB, Carrer de Can Magrans s/n, 08193 Cerdanyola del Vall\`es, Spain}
\def\addresso{DISPEA, Universit\`a di Urbino ``Carlo Bo'', Via S. Chiara, 27 61029 Urbino/INFN, Italy}
\def\addressp{European Space Operations Centre, European Space Agency, 64293 Darmstadt, Germany }
\def\addressq{Physik Institut, 
Universit\"at Z\"urich, Winterthurerstrasse 190, CH-8057 Z\"urich, Switzerland}
\def\addressr{SUPA, Institute for Gravitational Research, School of Physics and Astronomy, University of Glasgow, Glasgow, G12 8QQ, UK}
\def\addresss{Department d'Enginyeria Electr\`onica, Universitat Polit\`ecnica de Catalunya,  08034 Barcelona, Spain}
\def\addresst{Institut d'Estudis Espacials de Catalunya (IEEC), C/ Gran Capit\`a 2-4, 08034 Barcelona, Spain}
\def\addressu{Gravitational Astrophysics Lab, NASA Goddard Space Flight Center, 8800 Greenbelt Road, Greenbelt, MD 20771 USA}
\def\addressbb{Department of Mechanical and Aerospace Engineering, MAE-A, P.O. Box 116250, University of Florida, Gainesville, Florida 32611, USA}
\def\addresscc{Istituto di Fotonica e Nanotecnologie, CNR-Fondazione Bruno Kessler, I-38123 Povo, Trento, Italy}

\author{M~Armano}\affiliation{\addressh}
\author{H~Audley}\affiliation{\addressb}
\author{J~Baird}\affiliation{\addressc}
\author{P~Binetruy}\thanks{Deceased 30 March 2017}\affiliation{\addressc}
\author{M~Born}\affiliation{\addressb}
\author{D~Bortoluzzi}\affiliation{\addressf}
\author{E~Castelli}\affiliation{\addressi}
\author{A~Cavalleri}\affiliation{\addresscc}
\author{A~Cesarini}\affiliation{\addresso}
\author{A\,M~Cruise}\affiliation{\addressj}
\author{K~Danzmann}\affiliation{\addressb}
\author{M~de Deus Silva}\affiliation{\addressa}
\author{I~Diepholz}\affiliation{\addressb}
\author{G~Dixon}\affiliation{\addressj}
\author{R~Dolesi}\affiliation{\addressi}
\author{L~Ferraioli}\affiliation{\addressl}
\author{V~Ferroni}\affiliation{\addressi}
\author{E\,D~Fitzsimons}\affiliation{\addressm}
\author{M~Freschi}\affiliation{\addressa}
\author{L~Gesa}\affiliation{\addressn}
\author{F~Gibert}\affiliation{\addressi}
\author{D~Giardini}\affiliation{\addressl}
\author{R~Giusteri}\affiliation{\addressi}
\author{C~Grimani}\affiliation{\addresso}
\author{J~Grzymisch}\affiliation{\addressh}
\author{I~Harrison}\affiliation{\addressp}
\author{G~Heinzel}\affiliation{\addressb}
\author{M~Hewitson}\affiliation{\addressb}
\author{D~Hollington}\affiliation{\addressd}
\author{D~Hoyland}\affiliation{\addressj}
\author{M~Hueller}\affiliation{\addressi}
\author{H~Inchausp\'e}\affiliation{\addressc}\affiliation{\addressbb}\email[Corresponding author: ]{hinchauspe@ufl.edu}
\author{O~Jennrich}\affiliation{\addressh}
\author{P~Jetzer}\affiliation{\addressq}
\author{N~Karnesis}\affiliation{\addressc}
\author{B~Kaune}\affiliation{\addressb}
\author{N~Korsakova}\affiliation{\addressr}
\author{C\,J~Killow}\affiliation{\addressr}
\author{J\,A~Lobo}\thanks{Deceased 30 September 2012}\affiliation{\addressn}
\author{I~Lloro}\affiliation{\addressn}
\author{L~Liu}\affiliation{\addressi}
\author{J\,P~L\'opez-Zaragoza}\affiliation{\addressn}
\author{R~Maarschalkerweerd}\affiliation{\addressp}
\author{D~Mance}\affiliation{\addressl}
\author{N~Meshksar}\affiliation{\addressl}
\author{V~Mart\'{i}n}\affiliation{\addressn}
\author{L~Martin-Polo}\affiliation{\addressa}
\author{J~Martino}\affiliation{\addressc}
\author{F~Martin-Porqueras}\affiliation{\addressa}
\author{I~Mateos}\affiliation{\addressn}
\author{P\,W~McNamara}\affiliation{\addressh}
\author{J~Mendes}\affiliation{\addressp}
\author{L~Mendes}\affiliation{\addressa}
\author{M~Nofrarias}\affiliation{\addressn}
\author{S~Paczkowski}\affiliation{\addressb}
\author{M~Perreur-Lloyd}\affiliation{\addressr}
\author{A~Petiteau}\affiliation{\addressc}
\author{P~Pivato}\affiliation{\addressi}
\author{E~Plagnol}\affiliation{\addressc}\email[Corresponding author: ]{plagnol@apc.in2p3.fr}
\author{J~Ramos-Castro}\affiliation{\addresss}
\author{J~Reiche}\affiliation{\addressb}
\author{D\,I~Robertson}\affiliation{\addressr}
\author{F~Rivas}\affiliation{\addressn}
\author{G~Russano}\affiliation{\addressi}
\author{J~Slutsky}\affiliation{\addressu}
\author{C\,F~Sopuerta}\affiliation{\addressn}
\author{T~Sumner}\affiliation{\addressd}
\author{D~Texier}\affiliation{\addressa}
\author{J\,I~Thorpe}\affiliation{\addressu}
\author{D~Vetrugno}\affiliation{\addressi}
\author{S~Vitale}\affiliation{\addressi}
\author{G~Wanner}\affiliation{\addressb}
\author{H~Ward}\affiliation{\addressr}
\author{P\,J~Wass}\affiliation{\addressd}\affiliation{\addressbb}
\author{W\,J~Weber}\affiliation{\addressi}
\author{L~Wissel}\affiliation{\addressb}
\author{A~Wittchen}\affiliation{\addressb}
\author{P~Zweifel}\affiliation{\addressl}

\collaboration{LISA Pathfinder Collaboration}
\noaffiliation

\date{\today}
\makeatletter

% Search and replace example with OgreKit:
% Search: \\vec{(?<i>\w)}_\\indice{(?<o>\w/\w)}
% Replace: \\myvec{\g<i>}{\g<o>}

\makeatother

\begin{abstract}

The science operations of the LISA Pathfinder mission has  demonstrated the feasibility of sub-femto-g free-fall of macroscopic test masses necessary to build a LISA-like gravitational wave observatory in space. While the main focus of interest, i.e. the optical axis or the $x$-axis, has been extensively studied, it is also of interest to evaluate the stability of the spacecraft with respect to all the other degrees of freedom. The current paper is dedicated to such a study, with a focus set on an exhaustive and quantitative evaluation of the imperfections and dynamical effects that impact the stability with respect to its local geodesic. A model of the complete closed-loop system provides a comprehensive understanding of each part of the in-loop coordinates spectra. As will be presented, this model gives very good agreements with LISA Pathfinder flight data. It allows one to identify the physical noise source at the origin and the physical phenomena underlying the couplings. From this, the performances of the stability of the spacecraft, with respect to its geodesic, are extracted as a function of frequency. Close to  $1 mHz$, the stability of the spacecraft on the $X_{SC}$, $Y_{SC}$ and $Z_{SC}$  degrees of freedom is shown to be of the order of  $5.0\ 10^{-15} m\ s^{-2}/\sqrt{Hz}$  for X and $4.0 \ 10^{-14} m\ s^{-2}/\sqrt{Hz}$  for Y and Z. For the angular degrees of freedom, the values are of the order  $3\ 10^{-12} rad\ s^{-2}/\sqrt{Hz}$ for $\Theta_{SC}$  and $3\ 10^{-13} rad\ s^{-2}/\sqrt{Hz}$  for  $H_{SC}$ and $\Phi_{SC}$. Below $1 mHz$, however, the stability performances are observed to be significantly deteriorated, because of the important impact of the star tracker noise on the closed loop system. It is worth noting that LISA is expected to be spared from such concerns, essentially as {\it Differential Wave-front Sensing}, an attitude sensor system of much higher precision, will be utilized for attitude control.

%The coordinates spectra breakdown also allow to identify the parameters which are particularly of interest in order to extrapolate LISA Pathfinder performances to LISA performances expectations.

\end{abstract}


%\keywords{}

\maketitle
\makeatletter

% Search and replace example with OgreKit:
% Search: \\vec{(?<i>\w)}_\\indice{(?<o>\w/\w)}
% Replace: \\myvec{\g<i>}{\g<o>}

\makeatother

\addcontentsline{toc}{section}{General Introduction}
\section{General Introduction}
\label{section: General introduction}

The stability of a space platform, understood as a property of low noise acceleration of the platform with respect to the local geodesic, is a quality that is often searched for in order to satisfy the requirements of scientific observations or to perform tests of fundamental physics. Examples of such activity range from high precision geodesy, gravity field and gradient measurements (GRACE, GOCE), experimental test of gravitation (GP-B, MICROSCOPE) and gravitational wave astronomy (LISA Pathfinder, LISA). Using two quasi free-falling test masses ({\it TMs}),  LISA Pathfinder ({\it LPF}) \cite{anza_ltp_2005} has demonstrated remarkable properties related to its stability and recent publications (see \cite{armano_sub-femto-g_2016} and  \cite{armano_LPF_Ultimate_2017}) have presented the observed performance along the axis joining its test masses. However, the importance of the stability of the LISA Pathfinder platform is not limited to this axis, therefore in this paper we present results associated to its 6 degrees of freedom. In order to evaluate these performance, it is necessary not only to make use of the internal measurement of its sensors and actuators but also to deduce the \textit{true} motion of the spacecraft {\it S/C} impacted by the imperfections of the sensor and actuator systems. Beyond this, it is also necessary to evaluate the relative motion between the test masses and the platform due to internal forces whose manifestation is hidden from most sensors because the closed-loop control scheme nulls the measurement of in-loop sensors on {\it LISA Pathfinder}. In this publication we first introduce the configuration of the LISA Pathfinder  platform and then the closed-loop control scheme that allows the observed performance to be reached. In order to understand how these performance are reached, we introduce a simplified linear time invariant State Space model which allows extrapolation from in-loop sensor outputs in order to obtain needed physical quantities otherwise unobserved. An important example of such quantity is the actual low frequency relative displacement between the test masses and the {\it S/C}, driven by the sensors noise and masked by fundamental properties of the closed-loop systems (further details at section \myhyperref{The Stability of the Spacecraft}). We show that this model is capable of reproducing the observations of the sensors to within a few percent and can therefore be relied upon. The last section, before the conclusion, is devoted to summing up all the effects that allow the stability of LPF w.r.t. its local geodesic over the six degrees of freedom to be deduced.

\section{The LISA Pathfinder Platform}
\label{section: The LISA Pathfinder Platform}

LISA Pathfinder \cite{armano_lisa_2015} aims to demonstrate that it is technically possible to make inertial reference frames in space at the precision required by low-frequency gravitational waves astronomy. Indeed, in a LISA-like observatory design \cite{amaro-seoane_laser_2017}, one needs excellent references of inertia inside each satellites in order to differentiate between spurious accelerations of the apparatus from gravitational radiations, which both result in detected oscillatory variation of the arm lengths of the spacecraft constellation. The quality of free-fall achieved along the $X$-axis, the axis of main interest (i.e. representing axes along LISA arms),  has already significantly exceeded expectations \cite{armano_sub-femto-g_2016}. In addition to limiting stray forces acting directly on the TM, the {\it LPF} differential acceleration result required stringent and specific control of the TM-SC relative motion, to limit elastic "stiffness" coupling and possible cross-talk effects. Besides, post=processing {\it software} corrections from modeling of such {\it S/C}-to-{\it TM} acceleration couplings have been proven to be necessary and efficient in order to extract measurements of residual acceleration exerted on the {\it TMs} only (such as inertial forces, stiffness coupling, cross-talk corrections \cite{armano_sub-femto-g_2016}).

The control scheme required challenging technologies permitting high precision sensing and actuation in order to finely track and act on the three bodies and  keep them at their working point. {\it LISA Technology Package (LTP)}, the main payload of LISA Pathfinder, was built to demonstrate the required performance \cite{armano_lisa_2015} and includes high performance sensing and actuation subsystems. The {\it Gravitational Reference Sensor} ({\it GRS}) \cite{dolesi_gravitational_2003} includes the $1.93 \si{kg}$ {\it Au-Pt} test mass, surrounded by a conducting electrostatic shield with electrodes that are used for simultaneous capacitive position sensing and electrostatic force actuation of the {\it TM}. The {\it Optical Metrology System (OMS)} \cite{heinzel_ltp_2004} uses heterodyne interferometry for high precision test mass displacement measurements. Angular displacements are sensed through the {\it Differential Wavefront System (DWS)} using phase differences measured across the four quadrants of photodiodes. {\it Star Trackers (ST)} orients the spacecraft w.r.t. to a Galilean frame and  the {\it micro-thruster system}, a set of six Cold-Gas micro-thrusters (a technology already flown in space with ESA's GAIA mission \cite{Gaia_1} and CNES's Microscope mission \cite{microscope}), allows {\it S/C} displacement and attitude control along its six degrees of freedom. Note that LISA Pathfinder also has a NASA participation, contributing a set of 8 Colloidal Thrusters and the electronics/computer that control them \cite{anderson_experimental_2018}.

\begin{figure}[h]

\frame{\centerline{\includegraphics[scale=0.45, trim={0.0cm 0.5cm 1.3cm 0cm}, clip]{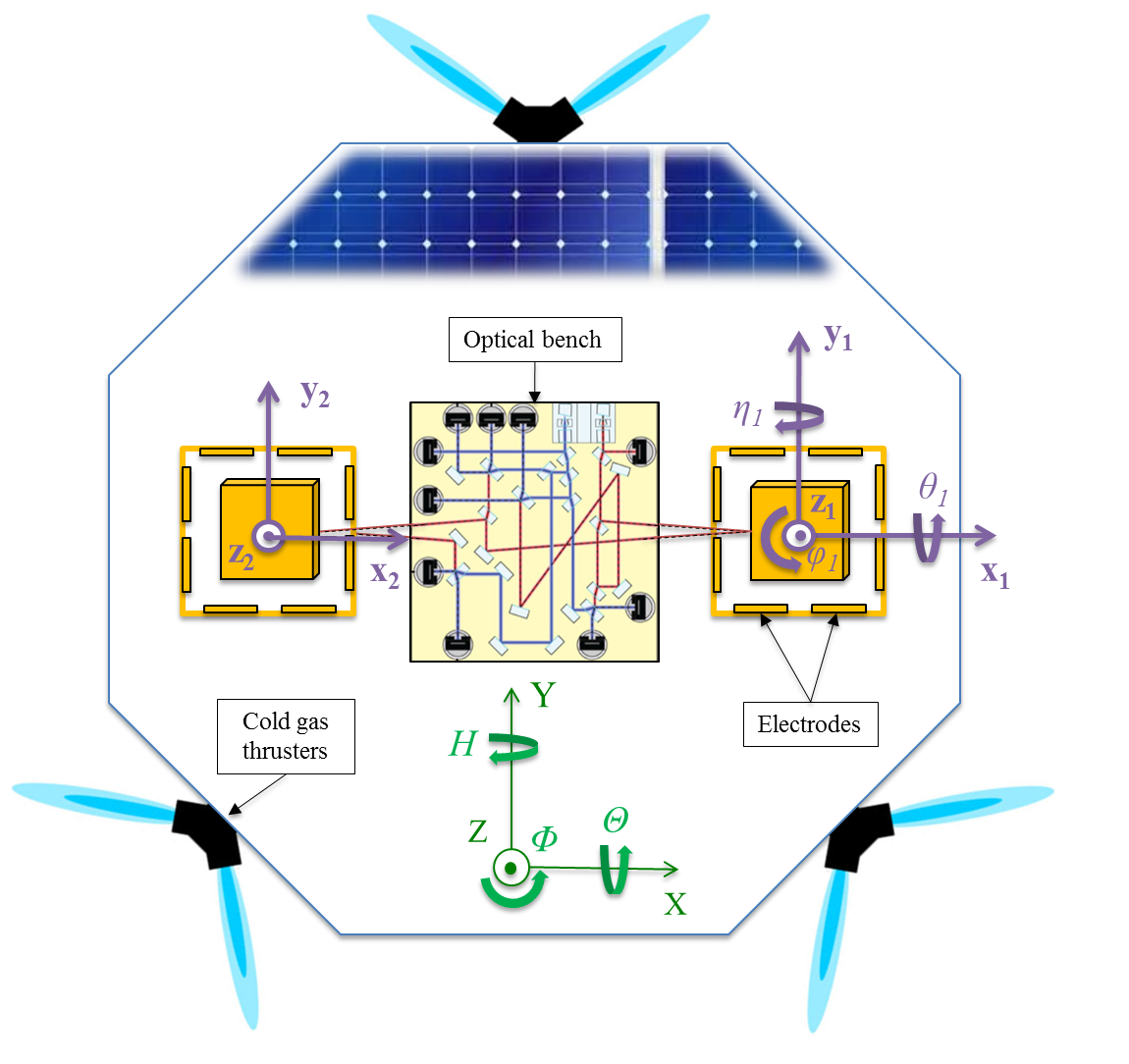}}}
\caption{Simplified sketch of LISA Pathfinder apparatus. The system of coordinates used to describe the displacement of the test masses (purple sets of axes) and of the spacecraft (green set of axes) are made explicit.}
\label{figure: LPF}

\end{figure}

\addcontentsline{toc}{section}{The Drag-Free and Attitude Control System (DFACS)}
\section{The Drag-Free and Attitude Control System (DFACS)}
\label{section: The DFACS}

The Drag-Free and Attitude Control System (DFACS) \cite{fichter_lisa_2005} is a central subsystem in LISA Pathfinder architecture. It has been designed by {\it Airbus Defence \& Space} \cite{Airbus}. It is devoted to achieve the control scheme that maintains the test masses to be free-falling at the centre of their electrode housings (translation control), to keep a precise alignment of the TMs w.r.t. the housing inner surfaces (rotation control) and to track the desired spacecraft orientation w.r.t. inertial frames (spacecraft attitude control). The translational control strategy is designed to limit any applied \textit{electrostatic suspension} forces on the {\it TMs} to the minimum necessary to compensate any \textit{differential} acceleration between the two {\it TMs}, while \textit{common mode} motion of these geodesic references, which essentially reflects {\it S/C} accelerations, is {\it drag-free} controlled with the micro-thrusters.  Limiting the applied actuation forces limits a critical acceleration noise from the actuator gain noise. This drag-free control is essentially used to counterbalance the  noisy motion of the spacecraft, which is both exposed to the space environment and largely to its own thrust noise. A linear combination of test mass coordinates inside their housing  along $Y-$ and $Z-$ axes are preferred for drag-free control, translational thrust being used to correct {\it common-mode} displacements while rotational actuations are performed to correct {\it differential-mode} displacements (see table \myhyperref{table: DFACS}, entries 5-8). Due to the geometrical configuration of the experiment (see figure \myhyperref{figure: LPF}), differential $x-$displacements of the TMs cannot be corrected by the drag-free control. In this case, it is necessary to apply control forces on one of the {\it TMs} along the $X$-axis. The strategy used is to leave TM1 in pure free-fall while the second mass is forced to follow the first, in order to keep the relative position of the masses constant at low frequencies. The amount of electrostatic force required to achieve this is measured and accounted for in the computation of the acceleration noise experienced by the masses \cite{armano_sub-femto-g_2016}. This control scheme is called {\it suspension control}. All the angular coordinates of the test masses (except rotation around $x_\indice{1}$) are controlled by the suspension control scheme (see table \myhyperref{table: DFACS}). The attitude of the spacecraft w.r.t. Galilean frame are supported by the {\it attitude control}. Because the commanded torques on the satellite are driven by the {\it drag-free control} of the differential linear displacement of the masses along $Y-$ and $Z-$ axes, as previously mentioned, the attitude control is realized indirectly. First, the attitude control demands differential forces on the masses according to information coming from the star trackers. Then, the drag-free loop takes the baton and corrects the induced differential displacement by requiring a rotation of the spacecraft, thus executing the rotation imposed by the star trackers.

\begin{table}%[H] add [H] placement to break table across pages
\scriptsize
\caption{Control scheme of LISA Pathfinder in {\it nominal science mode}. For each system's dynamical coordinate, the table lists by which subsystem it is sensed, its associated control type and the actuator used. $d$ is the distance between the TMs}
\label{table: DFACS}
\begin{ruledtabular}
\begin{tabular}{|c|c|c|c|c|}
\# & Coordinates & Sensor system & Control Mode & Actuation system \\
1 & $\Theta$ & ST & Attitude & GRS ($Tx_\indice{1}$) \\
2 & $H$ & ST & Attitude & GRS ($Fz_\indice{2} - Fz_\indice{1}$) \\
3 & $\Phi$ & ST & Attitude & GRS ($Fy_\indice{2} - Fy_\indice{1}$) \\
4 & $x_\indice{1}$ & IFO & Drag-Free & $\mu$-thrust ($\uparrow$ X-axis) \\
5 & $\frac{y_\indice{2} + y_\indice{1}}{2}$ & GRS & Drag-Free & $\mu$-thrust ($\uparrow$ Y-axis) \\
6 & $\frac{z_\indice{2} + z_\indice{1}}{2}$ & GRS & Drag-Free & $\mu$-thrust ($\uparrow$ Z-axis) \\
7 & $\frac{y_\indice{2} - y_\indice{1}}{d}$ & GRS & Drag-Free & $\mu$-thrust ($\circlearrowleft$ Z-axis) \\
8 & $\frac{z_\indice{2} - z_\indice{1}}{d}$ & GRS & Drag-Free & $\mu$-thrust ($\circlearrowleft$ Y-axis) \\
%$z_\indice{2} - z_\indice{1}$ & GRS & Drag-Free & $\mu$-thrust ($\circlearrowleft$ Y-axis) \\
9 & $\theta_\indice{1}$ & GRS & Drag-Free & $\mu$-thrust ($\circlearrowleft$ X-axis) \\
10 & $x_\indice{12}$ & IFO & Suspension & GRS ($Fx_\indice{2}$) \\
11 & $\eta_\indice{1}$ & IFO & Suspension & GRS ($Ty_\indice{1}$) \\
12 & $\phi_\indice{1}$ & IFO & Suspension & GRS ($Tz_\indice{1}$) \\
13 & $\theta_\indice{2}$ & GRS & Suspension & GRS ($Tx_\indice{2}$) \\
14 & $\eta_\indice{2}$ & IFO & Suspension & GRS ($Ty_\indice{2}$) \\
15 & $\phi_\indice{2}$ & IFO & Suspension & GRS ($Tz_\indice{2}$) \\
\end{tabular}
\end{ruledtabular}
\end{table}

\makeatletter

\makeatother

\addcontentsline{toc}{section}{Steady-State Performances: a Frequency Domain Analysis}
\section{Steady-State Performances: a Frequency Domain Analysis}
\label{section: Steady-state performances: a frequency domain analysis}

This study focuses on the {\it LISA Pathfinder} data during the measurement campaign  where very long  \textit{"noise only runs"} were operated in {\it nominal science mode}\footnote{The control loop mode when the science measurements were performed. Table \myhyperref{table: DFACS} details the control scheme for this mode.}: data collected in April 2016 and January 2017 are considered here. The \textit{"noise only run"} denomination means that the closed-loop is left to operate freely without injecting any excitation signal of any kind. The 15 in-loop measurement read-out, listed in table \myhyperref{table: DFACS}, are studied in the frequency domain. As in-loop measurements, they do not strictly reflect the dynamical state (i.e. the true displacements) of the {\it TMs} inside their housings, but represent the error signal of the control loop for each measurement channel, the working point being zero for all the degrees of freedom except for the {\it S/C} attitude. The $10 \si{\hertz}$ sampled ten day long data-sets are processed through Welch's modified periodogram method \cite{welch_use_1967} \cite{vitale_data_2014} to estimate variance-reduced power spectral densities of the measurement outputs, using 15 $50\%$ overlapping Blackmann-Harris windowed average segments.

\begin{figure}[h]

\centerline{\includegraphics[scale=0.55, trim={0.0cm 0.0cm 0.0cm 0.0cm}, clip]{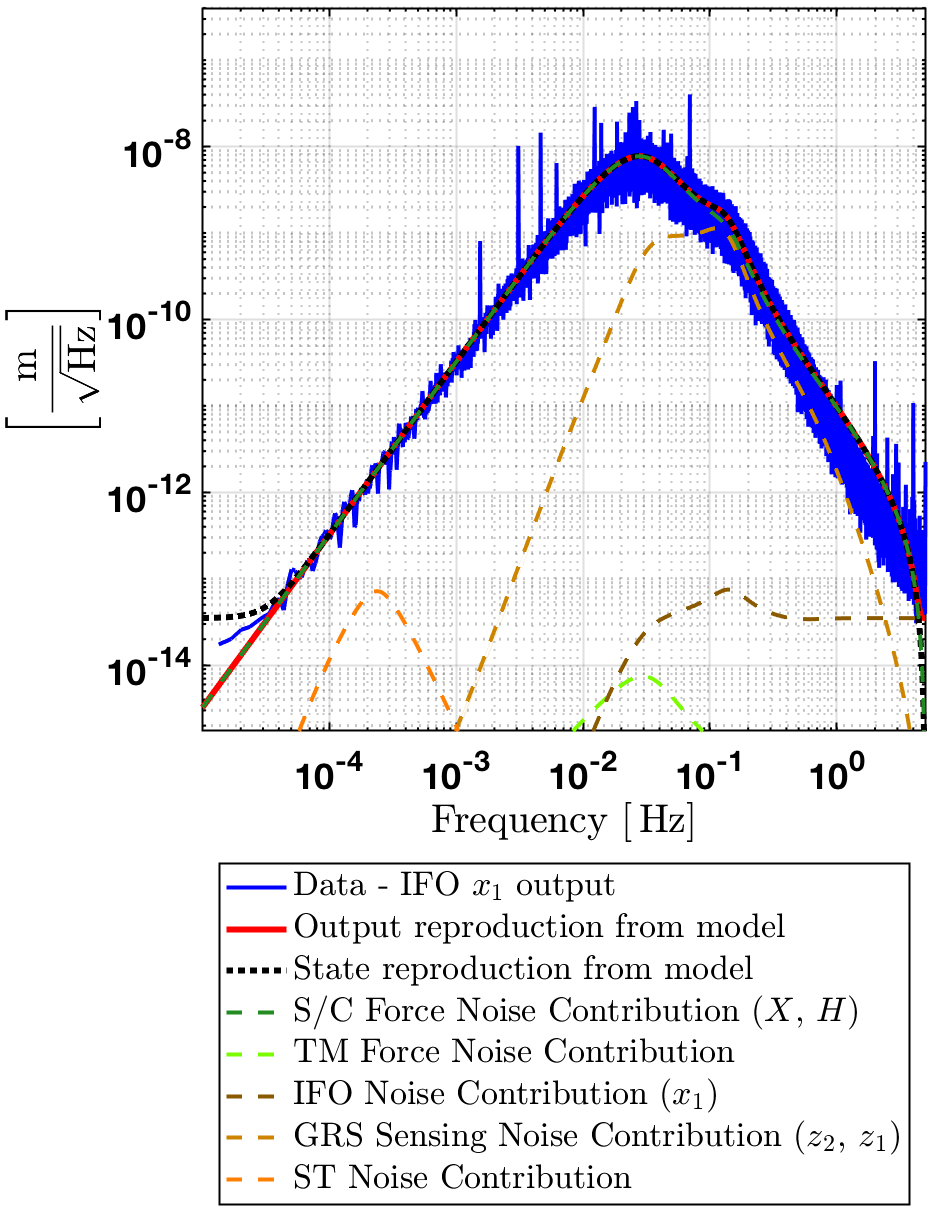}}
\caption{Decomposition of the in-loop optical measurement $o1$ spectrum, giving the linear motion of test mass 1 along x, into the various contributions from the sensing noise and the external disturbances. The blue curve corresponds to the {\it LISA Pathfinder} data. The red curve is built from modelled closed loop transfer functions (SSM model) with noise level from Table \myhyperref{table: MeasuredNoiseLevels}. The other curves give the individual contributions of the most relevant noise channels. \\
{\it This plot is assuming a {\it GRS} $z$  sensing with a noise floor of $1.8\ 10^{-9} \si{\metre\persqrthz}$ and a $\sfrac{1}{f}$ noise increase from $1 \si{\milli\hertz}$ and below. The {\it S/C} force and torque noise levels, extracted from equation \myhyperref{eq: ScThrusterNoiseForce},  are all measured to be consistent with white noise. The white noise level along the $X$ axis is measured to be of $0.17 \si{\micro\newton\persqrthz}$. See section \myhyperref{section: Sensing and actuation noise} and table \myhyperref{table: MeasuredNoiseLevels} for more details.}}
\label{figure: ControlPerformances_Measurements_DELAY_IFO_x1_EP}	

\end{figure}

As an example, figure \myhyperref{figure: ControlPerformances_Measurements_DELAY_IFO_x1_EP} shows the spectral density of the $o_1$ channel during the April 2016 run, i.e. the in-loop optical sensor read-out of the $x_1$ coordinate (cf. reference axes of figure \myhyperref{figure: LPF}). In the figure are traced together the observed data (in blue) and the sum of all the contributors (in red), as predicted by a state space model of the closed loop system \cite{weyrich_modelling_2008} (cf. section \ref{section: Spectrum decomposition using a State Space Model of the system} and equation \ref{equation: OutputsReconstruction}). 
The remaining lines show the break-down of the different components that contribute significantly to the sum: the external, out-of-loop forces applied on the {\it S/C} and the {\it GRS } sensing noise (mostly $z_1$ and $z_2$ sensing noise as visible after breaking down the contribution further) which are superimposed. The {\it S/C} force noise curve (dark green) is essentially due to the micro-thruster noise, for movements along X and rotation around Y. This has been demonstrated by using the {\it Colloidal} and cold-gas micronewton thruster systems alternatively and jointly \cite{anderson_experimental_2018}. Note that the presence of a strong {\it GRS} sensing noise component, around $ 0.1 \si{\hertz}$, is due to the control strategy. Further details about this model reconstruction, and other examples, are given in section \ref{section: Spectrum decomposition using a State Space Model of the system}.

The residual spectrum of $o_1$ reflects the frequency behaviour of the drag-free control gain. Below $ 0.1 \si{\hertz}$, the control loop gain is high and counters the noisy forces applied on the {\it S/C} (mostly thruster noise but also solar noise, etc.).

The drag-free gains continuously decrease with increasing frequencies to reach a minimum around $30 \si{\milli\hertz}$. The spectrum is conversely increasing as $f^2$, reaching its maximum jitter level of about $7 \si{\nano\meter\persqrthz}$ around $30 \si{\milli\hertz}$. At higher frequencies, the  ${f^{-2}}$ behaviour, due to the inertia of the {\it TMs}, is responsible for the spectrum drop. On the right end of the plot, above $1 \si{\hertz}$, one would normally see read-out noise floor only. However, as shown by the dark brown dashed line on the right end of the plot, the optical sensing noise is outstandingly low, less than $0.1 \si{\pico\meter\persqrthz}$ as already presented in \cite{armano_sub-femto-g_2016}, such that it has almost no perceptible impact on $o_1$ in the frequency domain of interest. The discrepancy between the data (blue line) and the model prediction (thick red line) visible above $0.5 \si{\hertz}$ is considered to be due to the imperfection of the $SSM$ model which does not reflect the non linear nature of the micronewton thruster system (e.g. pure delays). This has been confirmed by a comparison with ESA's simulator that does not assume this linear aspect.

%(see \hl{Appendix A} \myhyperref{Appendix: Appendix A}). 

% Note also that the $green$ dots represent the frequencies at which the differences between the model (thick red line) and the data (blue line) are minimised.

The spectrum breakdown also gives interesting information about the {\it Multiple-Input-Multiple-Output (MIMO)} nature of the in-loop dynamics. In figure \myhyperref{figure: ControlPerformances_Measurements_DELAY_IFO_x1_EP}, this appears clearly with the superimposed light brown dashed lines that shows the influence of the {\it GRS} sensing noise on the spectrum of $x_1$, yet sensed by the {\it OMS}. Indeed, $z_1$ and $z_2$ sensing noise induces noisy {\it S/C} $Y$-axis ($\eta$) rotations as expected from the control scheme exposed in table \myhyperref{table: DFACS}. This motion causes an apparent $x-$axis displacement of the {\it TM1} inside its housing, the projection depending on the geometrical position of the housing w.r.t. the centre of mass of the {\it S/C}. This effect competes with force noise on the {\it S/C} at high frequencies.

\addcontentsline{toc}{section}{Spectrum Decomposition Using a State Space Model of the System}
\section{Spectrum Decomposition Using a State Space Model of the System}
\label{section: Spectrum decomposition using a State Space Model of the system}

\begin{figure}[h]

\centerline{\includegraphics[scale=0.4, trim={1.0cm 3.5cm 1.0cm 2.5cm}, clip]{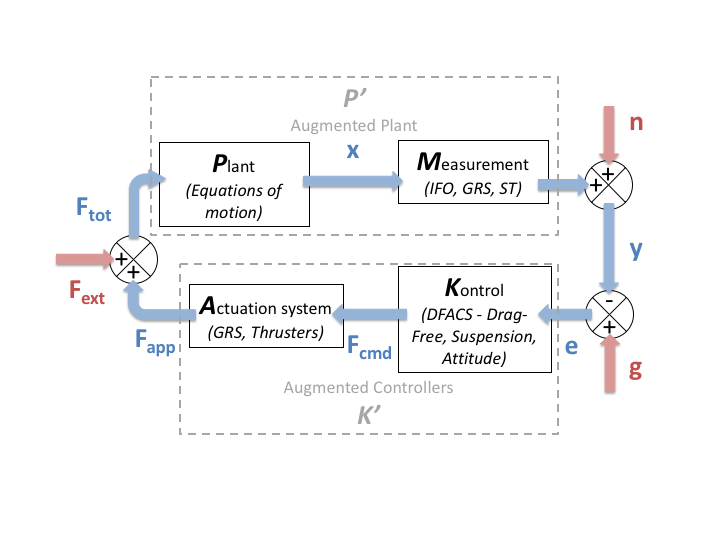}}
\caption{Simplified diagram of LISA Pathfinder closed loop system. The closed loop system is excited via three different {\it out-of-loop} signal channels: the guidance signal $g$ (null in the case of {\it "noise only run")}, the sensing noise $n$, and the external forces $F_{ext}$. The {\it in-loop} variables are indicated in blue, whereas the out-of-loop signals are in red. The {\it in-loop} variables are successively the states of motion $X$, the observed displacements $Y$, the error signal compared to targeted displacements $e$, the commanded $F_{cmd}$ forces, the effectively applied forces $F_{app}$ and the total resulting forces $F_{tot}$.}
\label{figure: ClosedLoop}

\end{figure}

As shown in the previous section, breaking down the data according to a physical model turns out to be very informative for tracking down the physical origins of the in-loop coordinates spectral behaviour. This model developed by the {\it LISA Pathfinder} collaboration \cite{weyrich_modelling_2008} (and implemented within the {\it LTPDA} software \cite{hewitson_data_2009}) is a {\it Linear Time Invariant (LTI)} {\it State Space Model (SSM)}, meaning that the modeled dynamical behaviour of the closed loop system does not depend on time, nor on the actual dynamical state. The latter is encoded within a state-space representation in such a way that the $N_{th}$-order differential system governing the dynamics transforms into a matrix system of $N$ first-order equations, thus benefiting from the matrix algebra arsenal. The linearity and stationarity of the model allows for straightforward conversions between time-domain {\it SSM} and frequency domain transfer functions. The superposition principle holds because of linearity and allows one to decompose all the resulting spectra into their various contributions by extracting the relevant {\it Single-Input-Single-Output (SISO)} transfer functions from the {\it MIMO} model. In that spirit, the in-loop sensor outputs can be decomposed with the help of closed loop transfer functions called {\it sensitivity functions}, which encode the sensitivity of the outputs to various out-of-loop disturbance signals, such as the sensing noise or the force noise applied on the bodies. Their respective transfer functions are named the $S$-gain and the $L$-gain as typically seen in the literature, and are given by the expressions:

\begin{align}
& S(f) = \frac{1}{1 + P'K'} & L(f) = \frac{P'}{1 + P'K'} \notag \\
&T(f) = \frac{P'K'}{1 + P'K'}
\label{equation: ClosedLoopTransferFunctions}
\end{align}

whereas $P$, standing for {\it Plant}, are the transfer functions of the dynamical system under {\it DFACS} control (forces/torques to displacements) and $K$ encodes the transfer functions of the {\it DFACS}. The prime symbols mention that these transfer functions are also including the transfer functions of the actuators ($K' = KA$) and of the sensors ($P' = MP$), for sake of notation simplification. Hence, these closed-loop transfer functions potentially depend on all the subsystem transfer functions involved in the control loop, and more significantly on the plant dynamics and the control laws. Figure \myhyperref{figure: ClosedLoop} gives an illustration of {\it LISA Pathfinder} closed-loop system which shows explicitly these transfer functions and the various {\it in-loop} and {\it out-of-loop variables}. Mathematically, the spectrum breakdown can be expressed in the following way:

\begin{equation}
\tilde{Y}^\exposant{q} = \sum\limits_\indice{p = x,y,z,\theta,\eta,\phi} L_\indice{\text{gain}}^\exposant{qp} \tilde{F}_\indice{\text{ext}}^\exposant{p} + \sum\limits_\indice{p = x,y,z,\theta,\eta,\phi} S_\indice{\text{gain}}^\exposant{qp} \tilde{n}^\exposant{p}
\label{equation: OutputsReconstruction}
\end{equation}

where $\tilde{Y}^\exposant{q}$, $\tilde{F}_\indice{\text{ext}}^\exposant{p}$ and $\tilde{n}^\exposant{p}$ are the Fourier transforms, for the degree of freedom $q$, of the associated in-loop sensor output, the {\it out-of-loop} force noises and sensing noises \cite{lpf_cl_analysis_symposium_zurich_2017} respectively. Because the $S$-gain (the sensitivity function) and the $L$-gain (load disturbance sensitivity function ) are both {\it MIMO} functions, a sum is performed over the extra dimension to account for cross-couplings effects that can have an important impact on the spectra (like the role played by {\it GRS} sensing noise on the spectra of the figure \myhyperref{figure: ControlPerformances_Measurements_DELAY_IFO_x1_EP}).

\begin{figure}[h]
\centerline{\includegraphics[scale=0.55, trim={0.15cm 0.0cm 0.0cm 0.0cm}, clip]{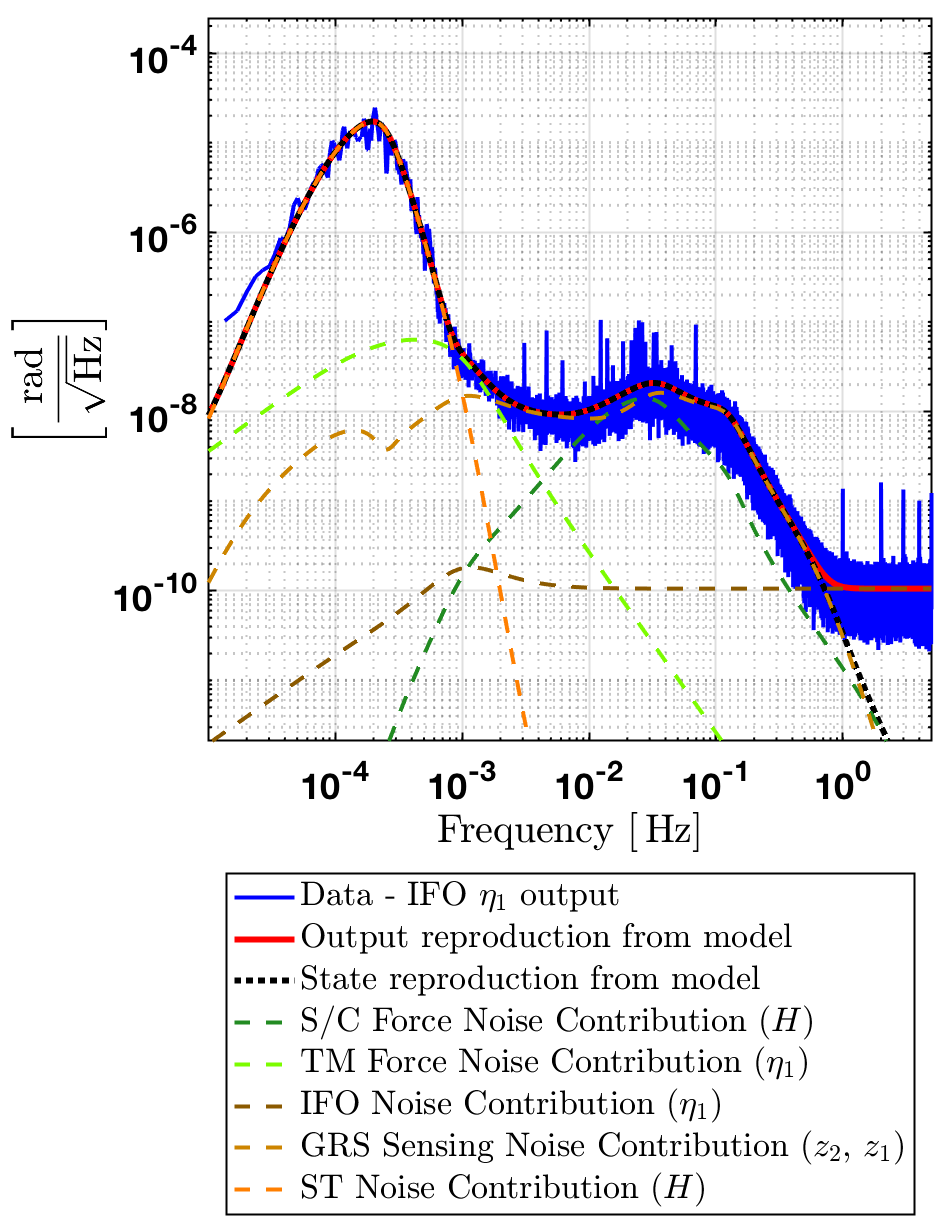}}
\caption{Decomposition of the in-loop DWS measurement $\eta_1$ spectrum into its various noise source contributions. The same explanations as in figure \ref{figure: ControlPerformances_Measurements_DELAY_IFO_x1_EP} apply. \\
{\it This plot is assuming: a {\it Star Tracker} sensing noise of $3.7\ 10^{-4} \si{\radian\persqrthz}$ at $0.2 \si{\milli\hertz}$, with a noise floor starting from $5 \si{\milli\hertz}$ at a level of $2.0\ 10^{-5} \si{\radian\persqrthz}$; a {\it TM} torque noise of $7.1\ 10^{-16} \si{\newton\persqrthz}$ across the whole bandwidth; a {\it GRS} $z$  sensing with a noise floor of $1.8\ 10^{-9} \si{\metre\persqrthz}$, a $\sfrac{1}{f}$ noise increase from $1 \si{\milli\hertz}$ and below, reaching a level of $4.6\ 10^{-9} \si{\metre\persqrthz}$ at $0.1 \si{\milli\hertz}$; a {\it S/C} torque noise around its $Y$ axis of $6.7\ 10^{-8} \si{\newton\metre\persqrthz}$; a white {\it DWS} $\eta$ noise of $1.05\ 10^{-10} \si{\radian\persqrthz}$. See section \myhyperref{section: Sensing and actuation noise} and table \myhyperref{table: MeasuredNoiseLevels} for more details.}}
\label{figure: ControlPerformances_Measurements_DELAY_IFO_eta1}
\end{figure}

A number of instances illustrate this type of decomposition. A case of particular interest is the spectra of the {\it TMs} angular displacements around $y$ and $z$ axes, which corresponding angles are labelled $\eta$ and $\phi$. The residuals behave in the most complex fashion because of numerous contributions that are equally competing to alter the TM orientation. As an example, figure \myhyperref{figure: ControlPerformances_Measurements_DELAY_IFO_eta1} shows the spectral density of the in-loop $\eta_{1}$ optical sensing output. Every control type of the {\it DFACS} - {\it drag-free}, {\it suspension} and {\it attitude} controls - has an influence on this plot. 

\begin{figure}[h!]
\centerline{\includegraphics[scale=0.55, trim={0.3cm 0.0cm 0.0cm 0.0cm}, clip]{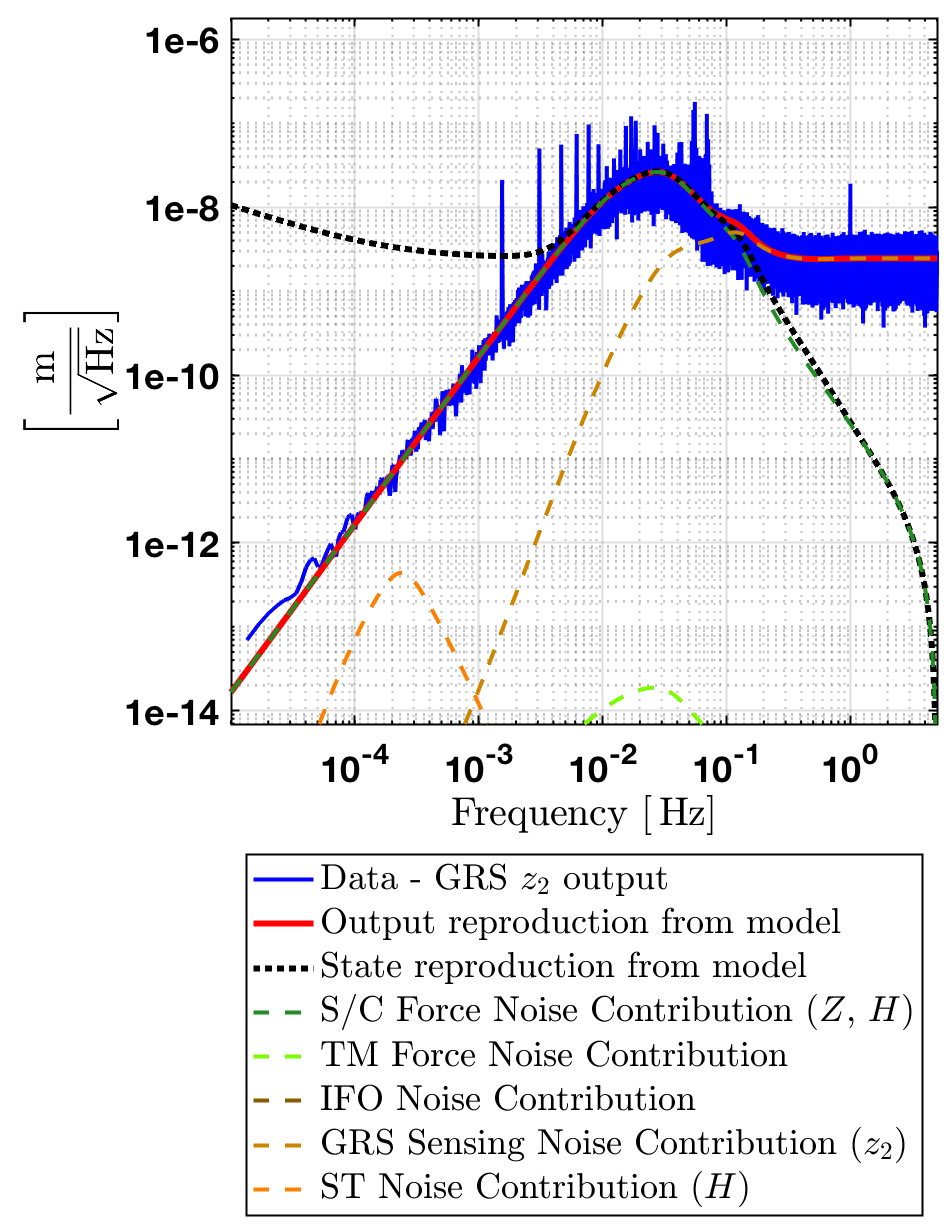}}
\caption{Decomposition of the in-loop {\it GRS} measurement $z_2$ spectrum into its various noise source contributions. The same explanations as in figure \ref{figure: ControlPerformances_Measurements_DELAY_IFO_x1_EP} apply. Note that the $z_2$ state (black dashed line) - or $z_2$ {\it "true motion"} - predicted from the model deviates very significantly from the {\it in-loop} {\it GRS} sensed $z_2$ displacement (data in blue, model in red). As discussed in detail at section \myhyperref{The Stability of the Spacecraft}, the {\it GRS} sensor noise induce a jitters of the {S/C} at low frequency (here $f < 1 \si{\milli\hertz}$) which is not observable from the in-loop {\it GRS} $z_2$ output. \\
{\it This plot is assuming a {\it GRS} $z$  sensing with a noise floor of $1.5\ 10^{-9} \si{\metre\persqrthz}$, a $\sfrac{1}{f}$ noise increase from $1 \si{\milli\hertz}$ and below, reaching a level of $4.5\ 10^{-9} \si{\metre\persqrthz}$ at $0.1 \si{\milli\hertz}$. The {\it S/C} force and torque noise levels, extracted from equation \myhyperref{eq: ScThrusterNoiseForce},  are all measured to be consistent with white noise. The white noise level along the $Z$ axis is measured to be of $4.6\ 10^{-7} \si{\newton\persqrthz}$. See section \myhyperref{section: Sensing and actuation noise} and table \myhyperref{table: MeasuredNoiseLevels} for more details.}}
\label{figure: ControlPerformances_Measurements_DELAY_IS_tm2_z}
\end{figure}

At the highest frequencies, the $\eta_1$ {\it DWS} sensing noise (dashed dark brown line) is the dominant factor. From $0.5 \si{\milli\hertz}$ to $0.5 \si{\hertz}$ there is a complex interplay between the external forces applied on the {\it S/C} (i.e. micronewton thrusters), residuals of the drag-free compensation, and the {\it GRS} sensing noise (light brown) of $y_1$, $y_2$ and $\theta_1$ that are all {\it Drag-Free} controlled.
At the lowest frequencies, the {\it Star Tacker} noise (orange line) which acts through the {\it Attitude Control} is the dominant source. It should be noticed that some of these noise sources, such as {\it GRS} $y_1$ and $y_2$ sensing noise (light brown), can be measured more directly through other channels; for example, see the analysis done in figure \myhyperref{figure: ControlPerformances_Measurements_DELAY_IS_tm2_z}. In the region around $1 \si{\milli\hertz}$, in figure \myhyperref{figure: ControlPerformances_Measurements_DELAY_IFO_eta1} back again, one is in presence of an ambiguity because observed spectrum can either be explained by the impact of {\it TM} torque noise of $\eta_1$ (light green) or by enhancing the reddening of the noise of $y_1$ and $y_2$ sensors. To remove this ambiguity, the reddening ($1/f$ behaviour) has been measured independently by \citet{Luigi_2017}.

Figure \myhyperref{figure: ControlPerformances_Measurements_DELAY_IS_tm2_z} shows the behaviour of the $z_1$ sensing output. This case is representative of what can also be observed for {\it Drag-Free} variables such as outputs $y_1$, $y_2$, $z_2$ and $\theta_1$. These spectra are much simpler than for the $\eta$ and $\phi$ channels. At the highest frequencies, the sensing noise (of $z_1$ in the figure) can be directly extracted. At lower frequencies, the behaviour of the spectra is essentially controlled by the $Z$ external forces exerted on the {\it S/C}, which means an estimation of this noise is also readily measurable.

\addcontentsline{toc}{section}{Sensing and Actuation Noise}
\section{Sensing and Actuation Noise}
\label{section: Sensing and actuation noise}

In the above section, equation \myhyperref{equation: OutputsReconstruction} shows how the out-of-loop force and sensing noises impact the observed spectra. In this section we begin by giving some illustrative examples of how the noise levels impact different frequency ranges and we then present our quantitative results in table  \myhyperref{table: MeasuredNoiseLevels}. Depending on the frequency range, a given noise will dominate the observed spectra. For instance, at the highest frequencies (typically from $f > 1 \si{\hertz}$), observed displacements of the {\it S/C} and the {\it TMs} are nearly insensitive to input external forces. Indeed, referring to equation \ref{equation: OutputsReconstruction}, $S_{\text{gain}} \approx 1$ and $L_{\text{gain}} \approx 0$ in such region, reflecting the fact the {\it inertia} of the bodies increases along with frequency, as a consequence of the $\sfrac{1}{f^{2}}$ behavior of the dynamical system (double integrator, i.e. from force to displacement). Consequently, the noise of a given sensor dominates the observed spectra in most cases, allowing for a straightforward determination of its level: see the dashed brown lines in figures \myhyperref{figure: ControlPerformances_Measurements_DELAY_IFO_eta1} or  \myhyperref{figure: ControlPerformances_Measurements_DELAY_IS_tm2_z} as examples. An exception is showed by figure \myhyperref{figure: ControlPerformances_Measurements_DELAY_IFO_x1_EP} where the sensor noise level is so low that even at the highest frequencies the sensitivity to the  sensing noise is not reached and a determination of its noise level cannot be made. In reference \citet{armano_LPF_Ultimate_2017}, the  x1 IFO noise level is indeed shown to be as small as  $30 \si{\femto\metre\persqrthz}$. At lower frequencies, below $5 \si{\milli\hertz}$, the out-of-loop {\it S/C} force noise usually dominates, see in figure \myhyperref{figure: ControlPerformances_Measurements_DELAY_IS_tm2_z} the dashed green line. The level of these noises can be determined with the help of the commanded forces or torques on/around the corresponding axis (see equation \ref{eq: ScThrusterNoise}, discussed later in the section). In the case of figure \myhyperref{figure: ControlPerformances_Measurements_DELAY_IFO_eta1} the situation is different and below $1 \si{\milli\hertz}$ the Star Tracker noise dominates (brown dashed line). We discuss later in this section how these noises are estimated.

In most instances, the frequency dependence of the noises are not directly measurable by an analysis of the spectra, because they are here not distinguishable from that of the closed-loop model transfer function. We used therefore the results of independent and dedicated investigations that were performed during the mission. For the capacitive sensing noises, we refer to \cite{Luigi_2017} which showed that these noises had a $\sfrac{1}{\sqrt{f}}$ (in amplitude) dependence below $1 \si{\milli\hertz}$ whereas, for the capacitive actuation noises, \cite{BillPaper} showed a $\sfrac{1}{f}$ dependence (in amplitude) below $1 \si{\milli\hertz}$. We have performed an analysis of all the observables ($x,\ y,\ z,\  \theta,\  \eta$ and  $\phi$) associated to {\it TM1} for a number of {\it noise only runs}. The results, obtained for April 2016 and January 2017, are collected in table \myhyperref{table: MeasuredNoiseLevels}. On the left panel of this table we list the sensing noises that we have used (see table \myhyperref{table: DFACS} for more details about the degrees of freedom). The first six lines correspond to the linear and angular sensing noises of {\it TM1}, whereas the last three lines correspond to the {\it S/C} Star Tracker noise. The third and fourth columns gives the values obtained in April 2016 and January 2017. The right panel of the table gives the values for the actuation noises. The first 6 lines correspond to the out-of-loop forces and torques on the {\it S/C} whereas the last 5 lines correspond to the capacitive actuation forces on {\it TM1}. The values given in this table correspond to the noise level at high frequencies. The $5th$ and $9th$ columns give the corner frequency and the power of the reddening of the noise below the corner frequency, when applicable. % {\color{blue} In this table, the notation $N.S.$ indicates that the spectra were not sensitive to the corresponding noise. $W.N.$ indicates that a white noise in frequency was assumed.}

Two special cases have to be highlighted. The last three observables on the left panel of the table (Star Tracker noises, entries {\it 7-9}) are obtained from a fit to the spectrum of the attitude control error signals out of the {\it DFACS}, corrected by the {\it S-gain} of the corresponding control loop. The attitude control is effective at frequencies well below the measurement bandwidth and the star tracker noise level dominates any actual {\it S/C} rotations in the latter bandwidth, which means that the attitude control error signal provides a direct measurement of the attitude sensor noise essentially, as confirmed by the state space model of the closed-loop system.  From these time series, a fit is obtained assuming a white noise at high frequencies, a rise for frequencies below $3 \si{\milli\hertz}$ and a saturation below $0.4 \si{\milli\hertz}$. The values given in table \myhyperref{table: MeasuredNoiseLevels} correspond to the white noise floor. It should be noted that the Star Tracker noises also show a number of features, i.e. peaks in the frequency domain around $5 \si{\milli\hertz}$, which are not included in the corresponding fits. With regards to the actuation noise on the right-hand panel of the table, the first six entries correspond to the noisy external forces and torques applied on the {\it S/C}, essentially by the thruster system itself (as discussed in section \ref{section: Steady-state performances: a frequency domain analysis}). This force noise can be measured from the calculation of the out-of-loop forces exerted on the {\it S/C}. Equations \ref{eq: ScThrusterNoise}, \ref{eq: ScThrusterNoiseForce} and \ref{eq: ScThrusterNoiseTorque} present such calculations, where the indices {\it ool} and {\it cmd} distinguish between out-of-loop forces and torques applied, and those commanded by the control loop (that oppose the {\it ool} forces and torques when the control gain is high). The meaning of the variables has been detailed in table \ref{table: DFACS}. The mass terms $m_\indice{\text{SC}}$, $m_\indice{1}$, $m_\indice{2}$, $I_\indice{\text{SC}}$, $I_\indice{1}$ are the mass and the inertia matrices of the {\it S/C} and of the two {\it TMs} respectively (the {\it TMs} are labelled by their numbers only). Also, $d$ denotes the distance between the working points of the two {\it TMs} (namely the centres of their housings).

\begin{equation}
\vec{n}^\exposant{\text{\text{thruster}}} \approx
\begin{bmatrix}
& \vec{F}_\indice{\text{SC}}^\exposant{\text{ool}} \\
& \vec{T}_\indice{\text{SC}}^\exposant{\text{ool}}
\end{bmatrix} \\
\label{eq: ScThrusterNoise}
\end{equation}

\begin{align}
\vec{F}_\indice{\text{SC}}^\exposant{\text{ool}} = m_\indice{\text{SC}}
\begin{bmatrix}
& \ddot{o_\indice{1}} \\
& \frac{\ddot{z_\indice{1}} + \ddot{z_\indice{2}}}{2} \\
& \frac{\ddot{y_\indice{1}} + \ddot{y_\indice{2}}}{2}
\end{bmatrix}
& - \vec{F}_\indice{\text{SC}}^\exposant{\text{cmd}}
\label{eq: ScThrusterNoiseForce}
\end{align}

\begin{align}
\vec{T}_\indice{\text{SC}}^\exposant{\text{ool}} = I_\indice{\text{SC}}
\begin{bmatrix}
& \ddot{\theta_\indice{1}} - \frac{T_\indice{x1}}{I_\indice{1, xx}} \\
& \frac{\ddot{z_\indice{2}} - \ddot{z_\indice{1}}}{2d} - \big( \frac{F_\indice{z2}}{m_\indice{2}} - \frac{F_\indice{z1}}{m_\indice{1}} \big) \\
& \frac{\ddot{y_\indice{1}} - \ddot{y_\indice{2}}}{2d} - \big( \frac{F_\indice{y1}}{m_\indice{1}} - \frac{F_\indice{y2}}{m_\indice{2}} \big)
\end{bmatrix}
& - \vec{T}_\indice{\text{SC}}^\exposant{\text{cmd}}
\label{eq: ScThrusterNoiseTorque}
\end{align}

The actuation noises of the capacitive actuators are addressed in the last five entries ({\it 7-11}) on the right-hand side of the table. The force noise for linear degrees of freedom ($y$ and $z$) are estimated from extrapolation of the $x$ noise. They are built from the addition of the Brownian noise level observed for the $x$ channels \cite{armano_sub-femto-g_2016} and a model-based extrapolation of the actuation noise for degrees of freedom other than $x$, which is expected to be dominant below $0.5 \si{\milli\hertz}$ because of the larger force and torque authorities along/around these axes. Regarding the torque noises (entries {\it 9-10}), their levels are measured at low frequency with the help of the following expression:

\begin{align}
\vec{T}_\indice{1/2}^\exposant{\text{ool}} = \frac{I_\indice{1/2}}{\sqrt{2}}
\begin{bmatrix}
& - \\
& \big( \ddot{\eta_\indice{2}} - \ddot{\eta_\indice{1}} \big) - \Big( \frac{T_\indice{y2}}{I_\indice{2, yy}} - \frac{T_\indice{y1}}{I_\indice{1, yy}} \Big) \\
& \big( \ddot{\phi_\indice{2}} - \ddot{\phi_\indice{1}} \big) - \Big( \frac{T_\indice{z2}}{I_\indice{2, zz}} - \frac{T_\indice{z1}}{I_\indice{1, zz}} \Big)
\end{bmatrix}
\label{eq: TmNoiseTorque}
\end{align}

In equation \myhyperref{eq: TmNoiseTorque}, calculating the difference between {\it IFO} angular displacement measurements of the two {\it TMs} rejects common mode noise angular accelerations of the {\it TMs}, therefore the impact of {\it S/C} to {\it TMs} angular acceleration. Subtracting capacitive commanded torques provides then an estimate of the out-of-loop torques on the {\it TMs}. However, this calculation is typically valid only below $1 \si{\milli\hertz}$, above which frequency sensing noise rapidly dominates. Below $1 \si{\milli\hertz}$, applying \myhyperref{eq: TmNoiseTorque} to the data, a flat noise torque is observed down to around $0.01 \si{\milli\hertz}$. As a conservative assumption, this white noise torque is averaged and extrapolated to the whole frequency band (hence labelled as white noise in table \myhyperref{table: MeasuredNoiseLevels}). Note that equation \myhyperref{eq: TmNoiseTorque} is not applicable to linear degrees of freedom $y$ and $z$, since their differential channels are essentially sensitive to the largely dominant {\it S/C} angular acceleration noise and are {\it Drag-Free} controlled. A similar limitation applies to the $\theta$ case.

In table \myhyperref{table: MeasuredNoiseLevels}, comparison between {\it April 2016} and {\it January 2017} data sets allows to appreciate the consistency between the {\it "noise runs"} and the stationnarity of the sensor and actuator performances. It is worth mentioning that independent studies by \cite{anderson_experimental_2018} also show consistent results and similar performances for the cold-gas thrusters at different times of the mission ({\it September 2016} and {\it April 2017}).

\begin{table*}[t]
\centering
\caption{Table of sensing and actuation noises. $N.S$ stands for "Not Sensitive" and ST for Star Tracker. $W.N$. stands for white noise. See the text for further explanations.}
\label{table: MeasuredNoiseLevels}
\begin{tabular}{|c||c|c|c|c||c|c|c|c|}
\hline
\# & \bf{Sensing Noise} & \bf{\hspace{1 mm}Apr. 2016\hspace{1 mm}} & \bf{\hspace{1 mm}Jan. 2017\hspace{1 mm}} & {$f_c : \alpha$} & \bf{Actuation Noise} & \bf{\hspace{1 mm}Apr. 2016\hspace{1 mm}} & \bf{\hspace{1 mm}Jan. 2017\hspace{1 mm}} & \bf{$f_c : \alpha$} \\ \hline

1 & $o_1 \ (\si{\metre\persqrthz})$ & $35.0\ 10^{-15}$ & $N.S$ &  $N.S$   & $Ext.\ X  \ (\si{\newton\persqrthz})$     &                                      $1.7\ 10^{-7}$    & $1.9\ 10^{-7}$            &  $W.N.$ \\ \hline

2 & $y_1 \ (\si{\metre\persqrthz})$ & $1.5\ 10^{-9}$ & $1.5\ 10^{-9}$   & $\hspace{1 mm}1\si{\milli\hertz}: -0.5\hspace{1 mm}$ & $Ext.\ Y  \ (\si{\newton\persqrthz})$      & $1.4\ 10^{-7}$   & $1.7\ 10^{-7}$            & $W.N.$  \\ \hline

3 & $z_1 \ (\si{\metre\persqrthz})$ & $1.8\ 10^{-9}$ & $1.7\ 10^{-9}$   & $1\si{\milli\hertz}: -0.5$ & $Ext.\ Z  \ (\si{\newton\persqrthz})$      & $4.6\ 10^{-7}$          & $3.3\ 10^{-7}$      & $W.N.$   \\ \hline

4 & $\theta_1 \ (\si{\radian\persqrthz})$     & $9.4\ 10^{-8}$ & $1.1\ 10^{-7}$ & $1\si{\milli\hertz}: -0.5$ &  $Ext.\ \theta \ (\si{\newton\metre\persqrthz})$      & $9.8\ 10^{-8}$     & $7.5\ 10^{-8}$    &  $W.N.$\\ \hline

5 & $\eta_1 \ (\si{\radian\persqrthz})$ & $1.05\ 10^{-10}$ & $1.06\ 10^{-10}$ &$1\si{\milli\hertz}: -0.5$ & $Ext.\  \eta\ (\si{\newton\metre\persqrthz})$      & $6.7\ 10^{-8}$ & $7.8\ 10^{-8}$ & $W.N.$ \\ \hline

6 & $\phi_1 \ (\si{\radian\persqrthz})$ & $2.05\ 10^{-10}$ & $2.05\ 10^{-10}$ &$1\si{\milli\hertz}: -0.5$ & $Ext.\  \phi\ (\si{\newton\metre\persqrthz})$ & $1.7\ 10^{-7}$ & $1.9\ 10^{-7}$ & $W.N.$ \\ \hline

7 & ST  $\theta \ (\si{\radian\persqrthz})$ & $2.1\ 10^{-6}$ &  $2.1\ 10^{-6}$ & {\it see text} &  $Fy_1 \ (\si{\newton\persqrthz})$      &  \multicolumn{2}{c|}{$8.0\ 10^{-15}$} & $1\si{\milli\hertz}: -1$ \\ \hline

8 & ST $\eta \ (\si{\radian\persqrthz})$ & $2.0\ 10^{-5}$ & $2.0\ 10^{-5}$ & {\it see text} &  $Fz_1 \ (\si{\newton\persqrthz})$      &  \multicolumn{2}{c|}{$8.0\ 10^{-15}$} &  $1\si{\milli\hertz}: -1$ \\ \hline

9 & ST  $\phi \ (\si{\radian\persqrthz})$ & $3.2\ 10^{-6}$ & $3.2\ 10^{-6}$ & {\it see text} &  $Ty_1 \ (\si{\newton\metre\persqrthz})$      &  $7.1\ 10^{-16}$ & $7.4\ 10^{-16}$            &  {\it W.N.} \\ \hline

10 &           &          &            &              &  $Tz_1 \ (\si{\newton\metre\persqrthz})$      & $2.6\ 10^{-16}$ & $1.9\ 10^{-16}$            &      {\it W.N.}        \\ \hline

% 11 &           &          &            &              &  $Tz_1 \ (\si{\femto\newton\metre\persqrthz})$     & \multicolumn{2}{c|}{$8.0\ 10^{-3}$}            &    $\hspace{1 mm}1\si{\milli\hertz}: -1\hspace{1 mm}$  \\ \hline
\end{tabular}
\end{table*}

\addcontentsline{toc}{section}{The Stability of the Spacecraft}
\section{The Stability of the Spacecraft}
\label{The Stability of the Spacecraft}

It has been shown in the previous sections that the {\it SSM} was able to reproduce and explain the in-loop observations of the linear and angular displacements of the {\it TMs} relative to the {\it S/C} through breaking down the control residuals into the respective contributions of the individual noise sources. This model can now be used to assess physical quantities that are out of reach of on-board sensors, such as "true displacement" of the bodies and their acceleration w.r.t. their local inertial frame.

% Having discussed how the different actuation and sensing noise impact the observed linear and angular observables, it is possible to use the SSM model to infer the  stability of the \textit{SC} with respect to its \textit{local geodesic}. This can be done in two steps. 

Indeed, using properties of the Space State Model, one can extract  the  \textit{true} movement of the {\it S/C} with respect to the {\it TMs}. This is done using the following formula:

\begin{equation}
\tilde{\textbf{X}}_\indice{\text{SC}}^\exposant{q} = \sum\limits_\indice{p = x,y,z,\theta,\eta,\phi} L_\indice{\text{gain}}^\exposant{qp} \tilde{F}_\indice{\text{ext}}^\exposant{p} + \sum\limits_\indice{p = x,y,z,\theta,\eta,\phi} T_\indice{\text{gain}}^\exposant{qp} \tilde{n}^\exposant{p}
\label{equation: StatesReconstruction}
\end{equation}

%\begin{equation}
%\tilde{T}^\exposant{q} = \sum\limits_\indice{p = x,y,z,\theta,\eta,\phi} T_\indice{\text{gain}}^\exposant{qp} \tilde{n}^\exposant{p}
%\label{equation: OutputsReconstruction}
%\end{equation}

where $\tilde{\textbf{X}}_\indice{\text{SC}}^\exposant{q}$ is the Fourier transform, for the degree of freedom $q$, of the associated {\it state variable}, or alternatively called the {\it true displacement} (i.e, not the observed displacement) of the {\it TMs} with respect to the {\it S/C}. $T_\indice{\text{gain}}^\exposant{qp}$ is commonly named the $T\ gain$, or the complementary sensitivity function  \cite{lpf_cl_analysis_symposium_zurich_2017}. The difference between the {\it true displacement} and the {\it observed displacement} is that the former is estimated without applying the sensing noise whereas the latter corresponds to the response of the sensor output, i.e. with its noise. It should be noted, however, that the noise of previous time steps have an impact on the {\it true displacement}.

% In that sense the \textit{true} displacement do NOT represent the displacements in the absence of noise but rather the measurement of the displacement at a given time, before the sensing noise is applied to them(\hl{Henri, please check this}).

This important distinction is a classical feature of in-loop variables of feedback systems. A closed-loop system will force the variable of interest to its assigned guidance value, generally zero. To do this, for example, it will apply a correcting force to the {\it S/C} that will not only compensate for any external disturbances, but will also be triggered by the noise of the corresponding position sensor, indistinguishable from true motion from the point of view of the {\it DFACS}. As a result, when the sensing noise is the leading component, the compensating force will make the {\it S/C} jitter in the aim of canceling out the observed sensing noise. Hence the \textit{state variable} will exhibit this movement whereas the sensor will show a value tending to its guidance at low frequency. Figure \myhyperref{figure:ZSC_ACC_Decomposed}, discussed further in the text, illustrates this for the $Z$ axis acceleration.

Assuming {\it TM1} follows a perfect geodesic, and following LPF's DFACS philosophy (see table \myhyperref{table: DFACS}), the stability of the {\it S/C} is defined by the dynamic variables shown in equation \ref{equation: ScStates}:

%As these expressions concern the movements of the {\it S/C} and of the \textit{Test Masses}, and following LPF's DFACS philosophy (see table \myhyperref{table: DFACS}), the stability of the {\it S/C} is defined by the following definitions:

% \begin{itemize}

% \item{$\ddot{X}_\indice{SC} = \ddot{x}_\indice{1}$}

% \item{$\ddot{Y}_\indice{SC} = \frac{\ddot{y}_\indice{1} + \ddot{y}_\indice{2}}{2}$}

% \item{$\ddot{Z}_\indice{SC} = \frac{\ddot{z}_\indice{1} + \ddot{z}_\indice{2}}{2}$}

% \item{$\ddot{\Theta}_\indice{SC} = \ddot{\theta}_\indice{1}$}

% \item{$\ddot{H}_\indice{SC} = \frac{\ddot{z}_\indice{2} - \ddot{z}_\indice{1}}{d} $}

% \item{$\Phi_\indice{SC} = \frac{\ddot{y}_\indice{2} - \ddot{y}_\indice{1}}{d} $}

% \end{itemize}

\begin{align}
& \ddot{X}_\indice{\text{SC}} = \ddot{x}_\indice{1} && \ddot{\Theta}_\indice{\text{SC}} = \ddot{\theta}_\indice{1} \nonumber \\
& \ddot{Y}_\indice{\text{SC}} = \frac{\ddot{y}_\indice{1} + \ddot{y}_\indice{2}}{2} && \ddot{H}_\indice{\text{SC}} = \frac{\ddot{z}_\indice{2} - \ddot{z}_\indice{1}}{d} \nonumber \\
& \ddot{Z}_\indice{\text{SC}} = \frac{\ddot{z}_\indice{1} + \ddot{z}_\indice{2}}{2} && \ddot{\Phi}_\indice{\text{SC}} = \frac{\ddot{y}_\indice{2} - \ddot{y}_\indice{1}}{d}
\label{equation: ScStates}
\end{align}

where $d$ is the distance between the two {\it TMs}.

However, the {\it TMs} cannot embody perfect local inertial frames, as they inevitably experience some stray forces, though of very low amplitude as previously demonstrated in \cite{armano_sub-femto-g_2016}. Hence, as a second step, it is necessary to draw an estimation of the {\it TMs} acceleration w.r.t. their local inertial frames, and add it up to the relative acceleration between the {\it TMs} and the {\it S/C} calculated at the previous step. Reference \cite{armano_sub-femto-g_2016} provides the acceleration noise floor due to brownian noise ($S_{0}^{1/2} = 5.6 \si{\femto\metre\per\second\squared\persqrthz}$, divided by $\sqrt{2}$ for the acceleration of a single {\it TM}), to which is added, in accordance to \cite{armano_LPF_Ultimate_2017}, a $1/f$ component starting from around $0.5 \si{\milli\hertz}$ and below. The nature of this noise is still unknown to this day, but has to be included in the analysis nevertheless.

\begin{figure}[h]
\centerline{\includegraphics[scale=0.3, trim={0.0cm 0.0cm 0.0cm 0.0cm}, clip]{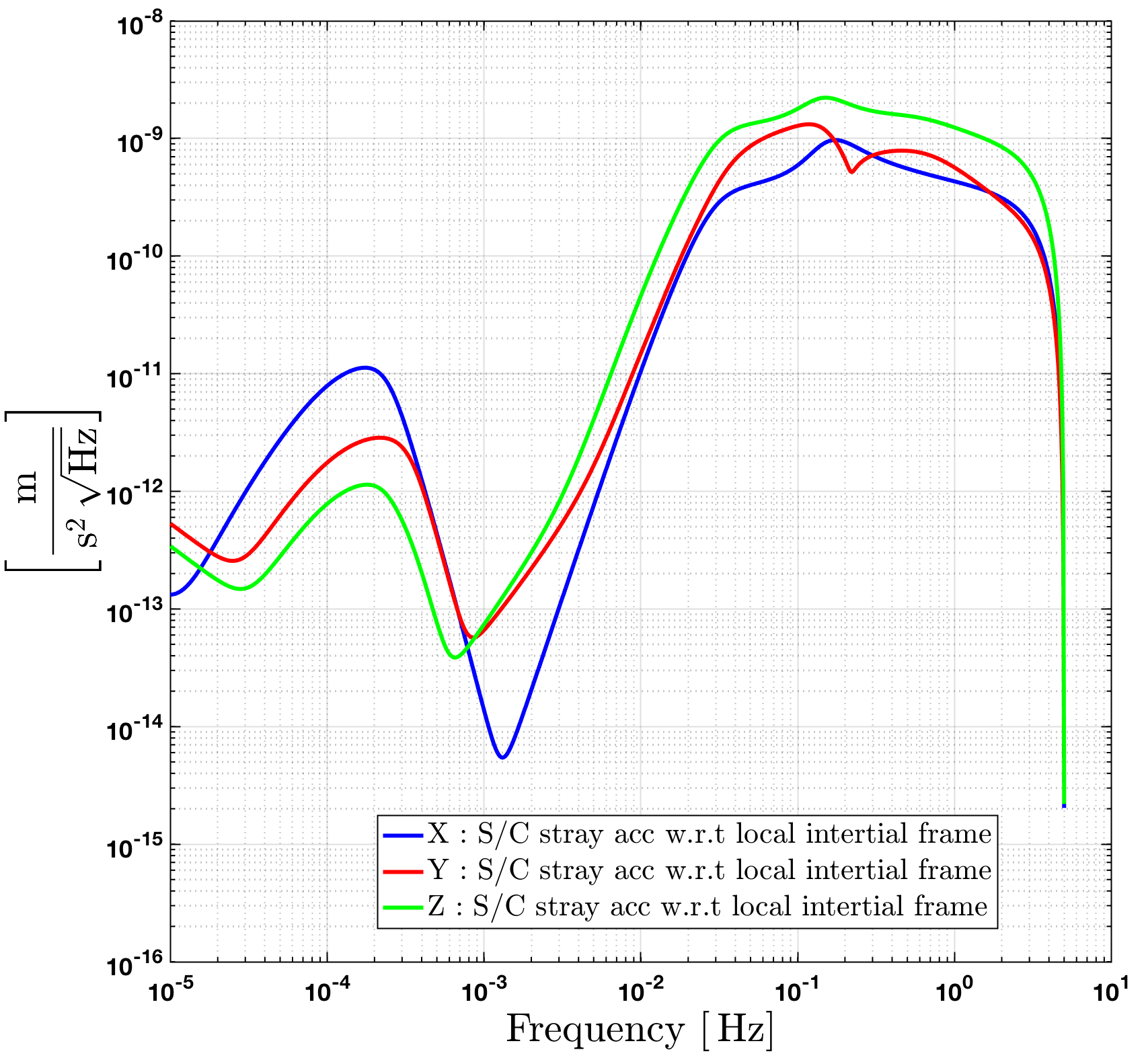}}
\caption{Stability of the {\it S/C} along X, Y and Z and as a function of frequency as simulated by the $LPF\ State\ Space\ Model$ using the parameters obtained from  the April 2016 \textit{"noise only run"}.}
\label{figure:stability_January2017_xyz}
\end{figure}

\begin{figure}[h]
\centerline{\includegraphics[scale=0.3, trim={0.0cm 0.0cm 0.0cm 0.0cm}, clip]{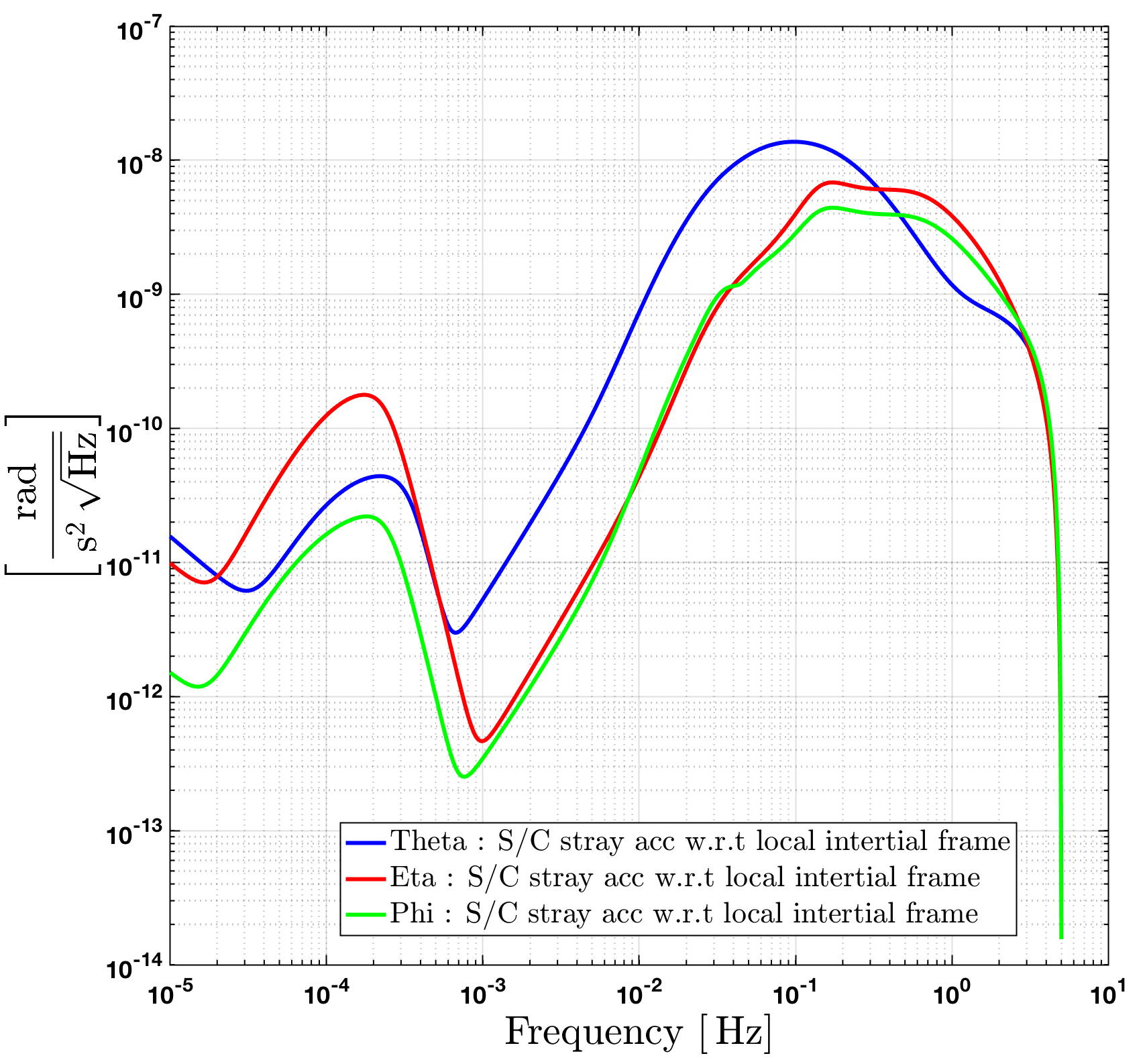}}
\caption{Stability of the {\it S/C} along $\theta$, $\eta$ and $\phi$ and as a function of frequency as simulated by the $LPF\ State\ Space\ Model$ using the parameters obtained from the April 2016 \textit{"noise only run"}.}
\label{figure:stability_January2017_angles}
\end{figure}

 Another factor that impacts the {\it LPF} stability is the {\it GRS} actuation noise. On the X axis, the impact is minimal because the actuation authority is set to a minimal value, just above the one required to compensate for the internal gravity gradient. On the other axes and on the angular degrees of freedom however, the actuation noise is expected to be dominant below $1 \si{\milli\hertz}$ according to model extrapolations for higher authority degrees of freedom \cite{BillPaper} (see table \myhyperref{table: MeasuredNoiseLevels} and discussion in section \myhyperref{section: Sensing and actuation noise}). Figures \myhyperref{figure:stability_January2017_xyz} and \myhyperref{figure:stability_January2017_angles}  show the stability (jitter) of the {\it S/C} and give a quantitative estimate of the \textit{true} movement of the {\it S/C} (for linear and angular degrees of freedom, respectively) relative to the local geodesic. In these figures at high frequencies ($ f \geqslant 2 \si{\hertz}$), one can note because of inertia of the {\it S/C}, the stability of the {\it S/C} improves with frequency. The region between $0.02 \si{\hertz}$ and $2 \si{\hertz}$ is explained by the characteristics of the control loop which incompletely compensate the noise of the micronewton thruster system. Below this frequency band, the accelerations are exponentially suppressed by the drag free loop until the effects of the capacitive sensing devices are observed, see for example around $2 \si{\milli\hertz}$ for the Y and Z accelerations in figure \myhyperref{figure:stability_January2017_xyz}. Note that this effect is not observed on the X axis because the optical sensing device has a very low noise, of the order of $35 \si{\femto\metre\persqrthz}$. Below $1 \si{\milli\hertz}$ the noise coming from the {\it Star Tracker} and the unexplained $1/f$ noise component explains the degradation of performance.

In figure \myhyperref{figure:stability_January2017_angles}, for $ 1 \si{\milli\hertz} \leqslant f \leqslant 0.1 \si{\hertz}$, one notices that $H$ and $\Phi$ stability are better that the one observed for $\theta$. This is because  $\theta$ is measured by electrostatic angular sensor of {\it TM1}, whereas $\eta$ and $\phi$ are measured by combinations of $z_1$ and $z_2$ and of $y_1$ and $y_2$ (see table \myhyperref{table: DFACS}), which have higher {\it SNR} benefiting from a larger lever arm between electrodes and noise averaging from electrodes redundancy compared to the single {\it GRS} angular channel.

\addcontentsline{toc}{section}{Decomposing the Stability of the Spacecraft}
\section{Decomposing the Stability of the Spacecraft}
\label{Decomposing the Stability of the Spacecraft}

\begin{figure}[h]
\centerline{\includegraphics[scale=0.47, trim={0.0cm 0.0cm 0.0cm 0.0cm}, clip]{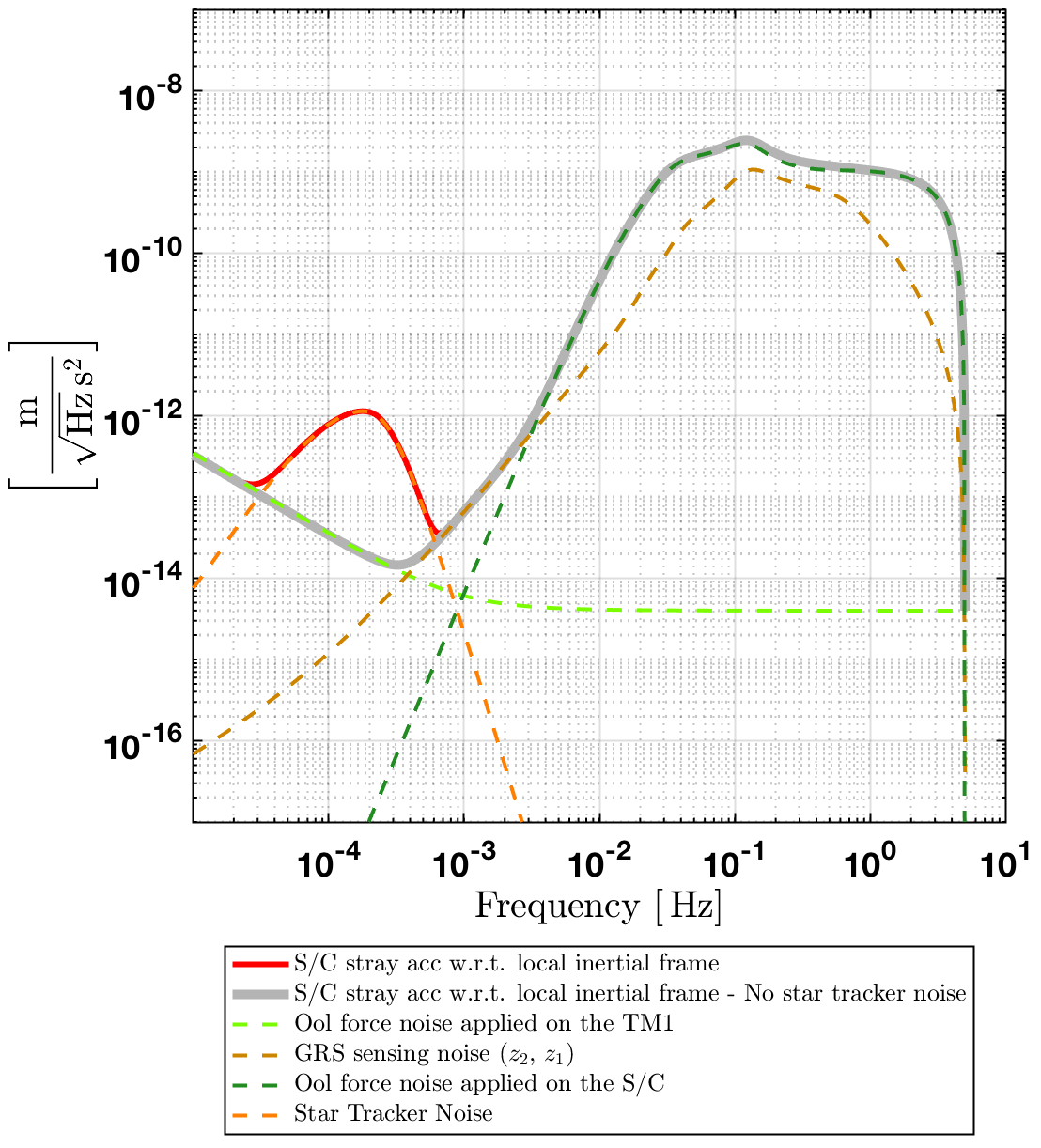}}
\caption{Decomposition of the stability of the {\it S/C} along the $Z$ axis as a function of frequency. The red lines shows the {\it SSM} prediction and the other lines the contribution to this. The main contributors are the {\it TMs} force noise (light green dashed line), the star tracker noise (orange dashed line), the GRS sensing noise (light brown dashed lines) and the micro-thruster noise (dark green dashed line). See the text for further explanation.}
\label{figure:ZSC_ACC_Decomposed}
\end{figure}

Figures \myhyperref{figure:stability_January2017_xyz} and \myhyperref{figure:stability_January2017_angles} present the stability of the {\it S/C} on all degrees of freedom. They show the complex behaviour of these stability performances. It is important to understand where the observed features come from. As an example, figure \myhyperref{figure:ZSC_ACC_Decomposed}  illustrates the decomposition of the acceleration stability on the $Z$ axis. Note that the $Z$ stability for LISA Pathfinder is calculated as the average $z$ values of {\it TM1} and of {\it TM2} (see equation \ref{equation: ScStates}). The red curve shows the sum of the listed contributions predicted by the {\it State Space Model}.

At the highest frequencies ($f \ >\ 0.5 \si{\hertz}$) the $Z$ sensing noise and the \textit{out of loop} noise (i.e. mainly thruster noises)  are predominant contributors. They are however countered by the inertia of the heavy {\it S/C} that does not allow it to move significantly, hence the roll-off of the red curve up to the Nyquist frequency at $f = 5 \si{\hertz}$ for this data. At lower frequencies ($5\ \si{\milli\hertz} \ <\ f\ <\  0.5 \ \si{\hertz}$) , the \textit{out of loop}  forces are attenuated by the control loops, hence the exponential decrease below $10 \si{\milli\hertz}$. Between $0.5 \si{\milli\hertz}$ and $5 \si{\milli\hertz}$, the {\it GRS} sensing noise on $Z$ is the dominant factor. This creates a movement of the {\it S/C} because the closed-loop system erroneously interprets this sensing noise as a non-zero position of the {\it TMs} to be corrected by the displacement of the {\it S/C}. Below this range the Star Tracker noise dominates, while at the lowest frequencies, the capacitive actuation noise governs the platform stability. Also indicated in this figure, and for illustration, is the effect of the Brownian noise which does not impact the stability performances on the $Z$ axis.

These explanations can be applied to all degrees of freedom with some differences for the $X$ axis. For this axis, the optical sensing noise is much smaller than {\it GRS} sensing noise and thus does not impact significantly the frequencies between $0.5 \si{\milli\hertz}$ and $5 \si{\milli\hertz}$. Another difference relates to the noise of capacitive actuation which is also much lower on X. At the lowest frequencies (around $0.01 \si{\milli\hertz}$), one observes the impact of the {\it "excess noise"} that is discussed in \cite{armano_LPF_Ultimate_2017}.

\addcontentsline{toc}{section}{The Impact of the Star Tracker Noise}
\section{The Impact of the Star Tracker Noise}
\label{The Impact of the Star Tracker Noise}

\begin{figure}[h]

\frame{\centerline{\includegraphics[scale=0.45, trim={0.0cm 0.5cm 1.3cm 2.0cm}, clip]{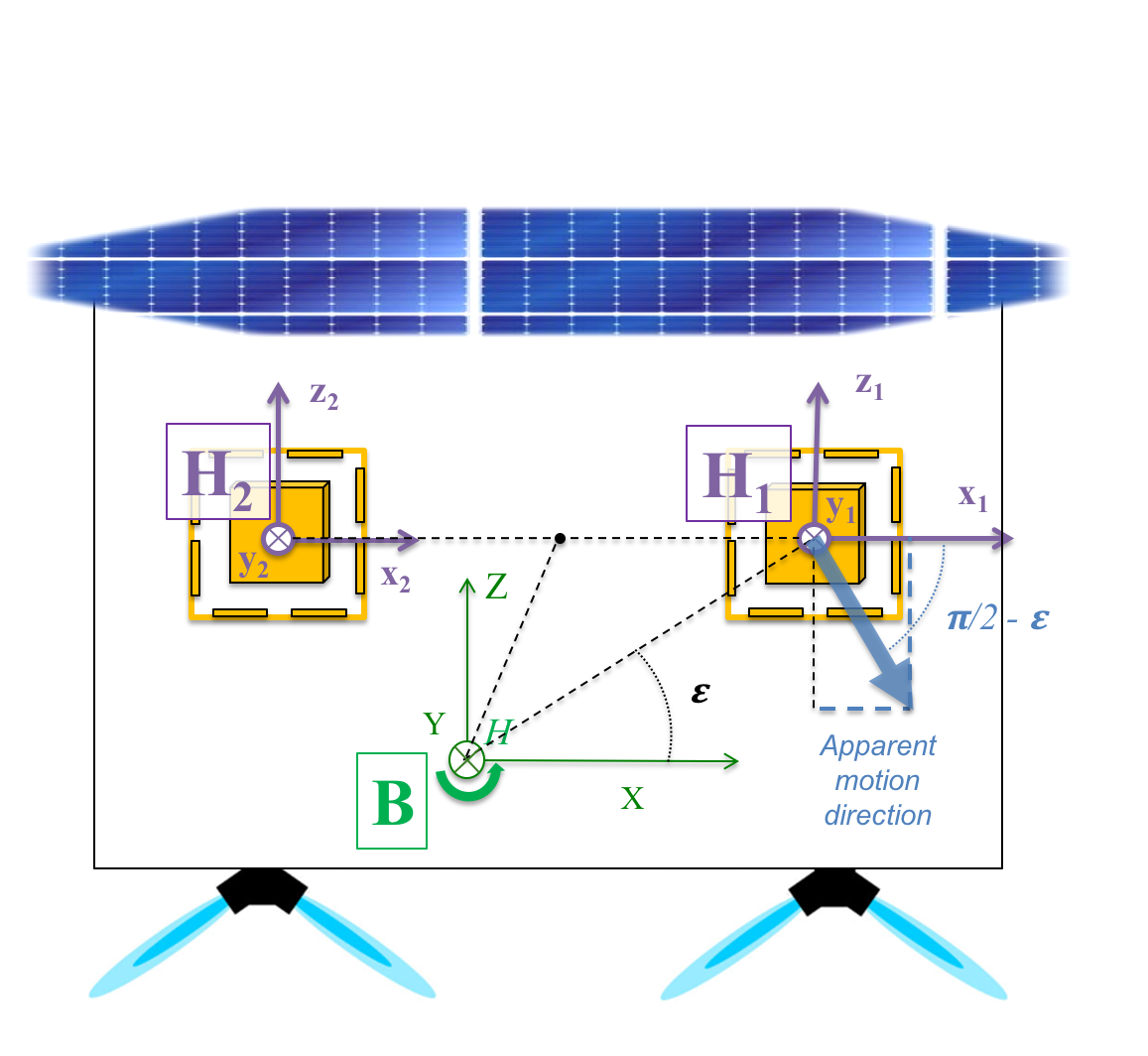}}}
\caption{Simplified sketch of {\it LISA Pathfinder} apparatus. The $xBz$ cross-section is represented here. The figure shows how any rotation of the {\it S/C}, in particular an $H$-rotation around the $Y$ axis, leads to apparent displacement of {\it TM1} inside its housing that has a significant component along the $X$-axis (proportional to $\sin{\epsilon}$ here) when the {\it S/C} center of mass is shifted from the center of the two housings.}
\label{figure: LPF_StNoiseProjection}

\end{figure}

Most of the contributions to {\it S/C} acceleration w.r.t. the local inertial observer are readily understandable. However, the impact of the Star Tracker noise is more subtle and needs explanation. The reason it impacts the platform stability is because the center of mass of the {\it S/C} does not coincide with the middle of the TM housing positions. By construction the center of mass is situated $6.25 \si{\centi\metre}$ below the housings along Z axis, but due to mechanical imperfections, it is also offset by a few millimeters  on the X and Y axis (see equation \ref{eq: CoM} and figure \ref{figure: LPF_StNoiseProjection}).

% EOM_SC_COM_X=0.005, EOM_SC_COM_Y=0.006, EOM_SC_COM_Z=0.47
% r_H1_B = 0.183   -0.006  0.0625
\begin{align}
\overrightarrow{BH_\indice{1}} =
\begin{bmatrix}
0.183 \\
-0.006 \\
0.0625
\end{bmatrix}
\si{\metre}
&&
\overrightarrow{BH_\indice{2}} =
\begin{bmatrix}
-0.193 \\
-0.006 \\
0.0625
\end{bmatrix}
\si{\metre}
\label{eq: CoM}
\end{align}

Because of this, {\it S/C} rotation jitter driven by the noisy star tracker sensor induces an apparent linear displacement of the {\it TMs} inside their housings. Such linear displacement has a significant component along X if the center of mass happens to be off-centered w.r.t. the middle of the line joining the two {\it TMs}. The projection of the force on X indeed scales with the sine of the angle $\epsilon$ made by the line joining the center of the housing and the {\it S/C} center of mass (that is to say the vector $\vec{BH_\indice{1}}$), and the axis joining the two housings (the vector $\vec{H_\indice{1}H_\indice{2}}$). Such an effect can more formally be interpreted as the result of the (so-called) {\it Euler force}, an inertial force proportional to {\it S/C} angular acceleration arising from the point of view of a non-inertial platform. Consequently, the {\it Drag-Free} control will react on and correct the (so-induced) displacement of {\it TM1} inside its housing. What was only an apparent force applied on the test mass then becomes a true force applied on the {\it S/C} along X through the micronewton thrusters and the feedback control. In fact, everything happens as though there existed a rotation-to-translation coupling of the {\it S/C} displacement, due to {\it S/C} geometry and DFACS activity. It is also worth noting that the impact on X-axis stability is observed to be greatly reduced in the case where the center of mass lies in the line joining TM housing centers.

Equation \ref{eq: EulerForces} provides an expression for the inertial forces responsible for the {\it TMs} displacement and equation \ref{eq: EulerForces_DfCorr} shows the {\it Drag-Free} control forces commanded to the micro-propulsion system in order to correct for the effect of the inertial forces. In these two equations, only the linear accelerations of the {\it S/C} are considered to emphasize the {\it rotation-to-translation} coupling of the {\it S/C} dynamics.

\begin{align}
\vec{a}^\exposant{\text{ST}} =
\begin{bmatrix}
a_\indice{x}^\exposant{\text{ST}} \\
a_\indice{y}^\exposant{\text{ST}} \\
a_\indice{z}^\exposant{\text{ST}}
\end{bmatrix}
=
\begin{bmatrix}
\Big[ \dot{\vec{\omega}}\times\overrightarrow{BH_\indice{1}} \Big]\cdot\hat{X} \\
\sfrac{1}{2}\Big[ \dot{\vec{\omega}}\times(\overrightarrow{BH_\indice{1}} + \overrightarrow{BH_\indice{2}}) \Big]\cdot\hat{Y} \\
\sfrac{1}{2}\Big[ \dot{\vec{\omega}}\times(\overrightarrow{BH_\indice{1}} + \overrightarrow{BH_\indice{2}}) \Big]\cdot\hat{Z}
\end{bmatrix}
\label{eq: EulerForces}
\end{align}

\begin{align}
\vec{F}_\indice{\text{DF}}^\exposant{\text{ST}} =
\begin{bmatrix}
F_\indice{\text{\text{DF, }}x}^\exposant{\text{ST}} \\
F_\indice{\text{DF, }y}^\exposant{\text{ST}} \\
F_\indice{\text{DF, }z}^\exposant{\text{ST}}
\end{bmatrix}
= - m_\indice{S/C}
\begin{bmatrix}
a_\indice{x}^\exposant{\text{ST}} \\
a_\indice{y}^\exposant{\text{ST}} \\
a_\indice{z}^\exposant{\text{ST}}
\end{bmatrix}
\label{eq: EulerForces_DfCorr}
\end{align}

\begin{figure}[h]
\centerline{\includegraphics[scale=0.47, trim={0.0cm 0.0cm 0.3cm 0.0cm}, clip]{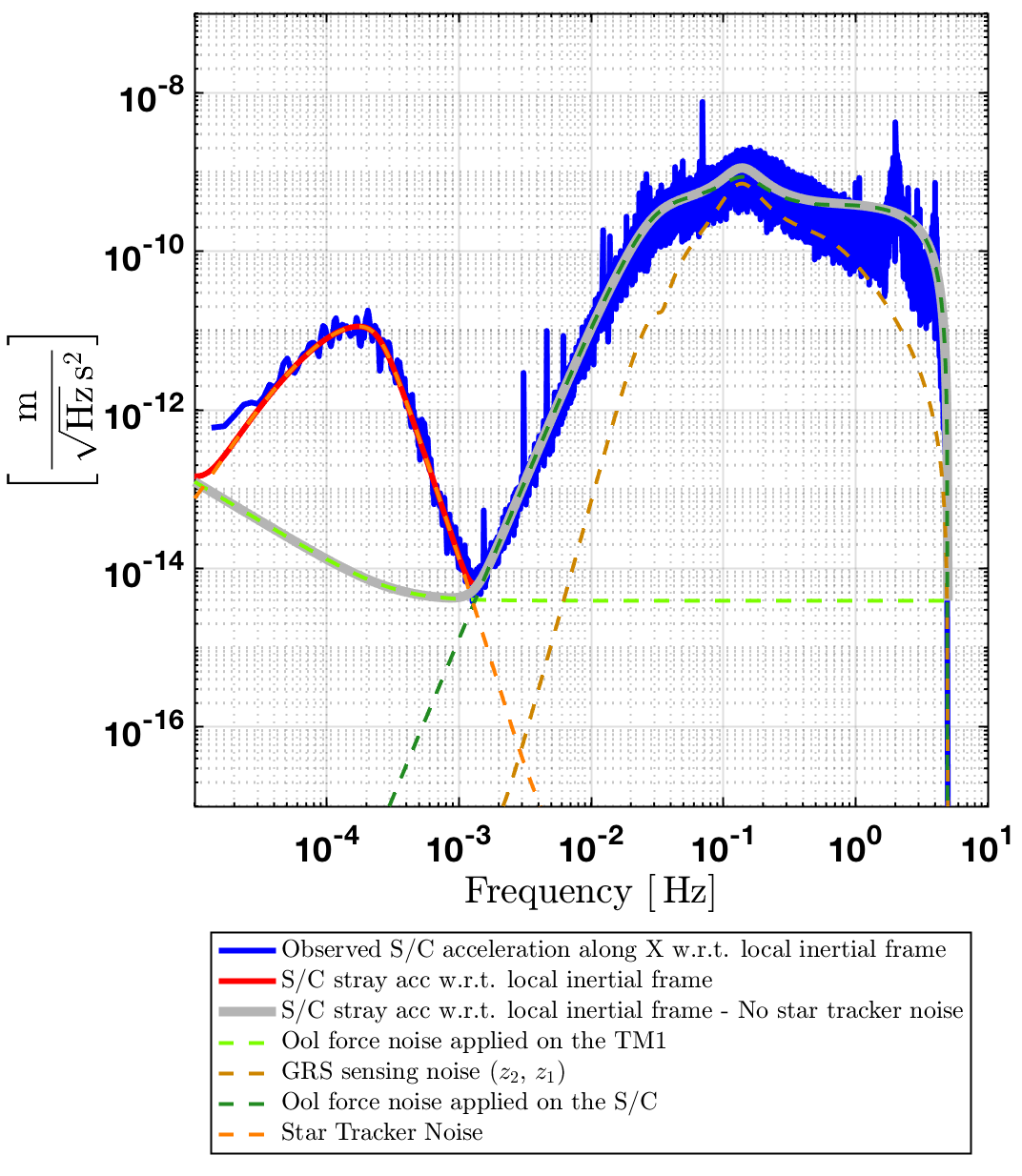}}
\caption{Stability of the {\it S/C} along $X$ as a function of frequency as simulated by the $LPF\ State\ Space\ Model$ {\it (red)} and as measured with a combination of observed data {\it (blue)} that takes into account the impact of the star tracker noise (according to equations \ref{eq: EulerForces} and \ref{eq: EulerForces_DfCorr}). Similar to the red line, the gray solid trace gives the model prediction for X stability of the S/C, though excluding the impact of the star tracker, hence providing a projection to the LISA observatory case (for which attitude control will be driven by {\it DWS} sensing of the inter-spacecraft laser beam, seen as an inertial attitude reference). The dashed-line traces show the model decomposition. This figure uses parameters and data obtained from the April 2016 \textit{"noise only run"}.}
\label{figure:ScStability_X_StNoiseImpact}
\end{figure}

The {\it State Space Model} predicts such indirect influence of the star tracker if set with a center of mass located off the axis joining the two {\it TMs}. The set values in the model are the ones shown in the equation \ref{eq: CoM}. Figure \ref{figure:ScStability_X_StNoiseImpact} shows the impact of the star tracker noise on the {\it S/C} stability along the X axis, together with all the other contributors already discussed in section \ref{Decomposing the Stability of the Spacecraft}. The blue trace is the combination of data sensor outputs given by equations \ref{eq: ScAccX_FromData} and \ref{eq: AngAcc}, and involving the double derivative of {\it TM1} interferometer readout $\ddot{o}_\indice{1}$ and the measurement of the force applied to the {\it S/C} along X to counteract the Euler force, presented in equation \ref{eq: EulerForces_DfCorr}. The angular acceleration of the satellite needed to compute the Euler force amplitude is recovered from {\it GRS} $\theta_\indice{1}$ measurements, and $z$ and $y$ differential measurements differentiated twice and corrected from the direct electrostatic actuation applied on the {\it TMs} $T_\indice{X}^\exposant{\text{cmd}}$, $F_\indice{z}^\exposant{\text{cmd}}$, $F_\indice{y}^\exposant{\text{cmd}}$ in order to trigger the {\it S/C} rotation according to the {\it DFACS} control scheme (see table \ref{table: DFACS}).

Figure \ref{figure:ScStability_X_StNoiseImpact} shows solid agreement between {\it SSM} predictions and computations from observations for the star tracker noise influence on stability along the X axis. It is visible in this figure that the star tracker noise deteriorated significantly platform stability at low frequencies by up to $3$ orders of magnitude at $0.1 \si{\milli\hertz}$. It is particularly noteworthy along the $X$ axis where high sensitivity of the optical sensor should have allowed for stability of the platform at the same level of quietness as the test mass itself (see the dark green dashed line in figure \ref{figure:ScStability_X_StNoiseImpact}), if it was not for the presence of a noisy sensor such as the star tracker (relatively to the other sensors of very high performance) within the DFACS loop. It is also worth noting that such stability performance decrease due to {\it S/C} attitude sensing noise will be largely mitigated in the case of LISA, where {\it Differential Wave-front Sensing} of the inter-spacecraft laser link will provide attitude measurement of much higher precision. Figure \ref{figure:ScStability_X_StNoiseImpact} shows a projection to {\it LISA} performances (light gray) following this consideration, hence excluding the contribution from the star tracker noise. Besides, in the case of LISA, studying the stability of the {\it S/C} center of mass is less relevant than studying the stability of the optical benches, which are geometrically much closer to the {\it TMs}, and thus less affected by the {\it rotation-to-translation} coupling here discussed.

\begin{equation}
a_\indice{X}^\exposant{\text{S/C, meas}} = -\ddot{o}_\indice{1} + \Big[ \dot{\vec{\omega}}^\exposant{\text{meas}}\times\vec{BH_\indice{1}} \Big]\cdot\hat{X}
\label{eq: ScAccX_FromData}
\end{equation}

\begin{align}
\dot{\vec{\omega}}^\exposant{\text{meas}} = 
\begin{bmatrix}
\dfrac{T_\indice{X}^\exposant{\text{cmd}}}{I_\indice{XX}} - \ddot{\theta_\indice{1}} \\
\dfrac{\sfrac{F_\indice{z2}^\exposant{\text{cmd}}}{m_\indice{2}} - \sfrac{F_\indice{z1}^\exposant{\text{cmd}}}{m_\indice{1}}}{H_\indice{1}H_\indice{2}} - \dfrac{\big(\ddot{z_\indice{2}} - \ddot{z_\indice{1}}\big)}{H_\indice{1}H_\indice{2}} \\
\dfrac{\sfrac{F_\indice{y2}^\exposant{\text{cmd}}}{m_\indice{2}} - \sfrac{F_\indice{y1}^\exposant{\text{cmd}}}{m_\indice{1}}}{H_\indice{1}H_\indice{2}} - \dfrac{\big(\ddot{y_\indice{2}} - \ddot{y_\indice{1}}\big)}{H_\indice{1}H_\indice{2}}
\end{bmatrix}
\label{eq: AngAcc}
\end{align}

%The dashed lines show the decomposition of this prediction, i.e. the thruster, sensing and other noise terms.  The cyan line shows the impact of the Brownian  and of the \textit{mystery} noise. The green line represents the impact of the \textit{GRS}  actuation noise which dominates at low frequencies. The blue line represent the data as seen by the \textit{GRS} sensor. The deviation at high frequency (($ f \geqslant 0.1\ \si{\hertz}$)) is due to the \textit{GRS} sensing noise. It does not represent a true motion of the \textit{SC} because, at these frequencies, the inertia of the \textit{SC} counters the applied thruster forces (\hl{TBC}).

\makeatletter

% Search and replace example with OgreKit:
% Search: \\vec{(?<i>\w)}_\\indice{(?<o>\w/\w)}
% Replace: \\myvec{\g<i>}{\g<o>}

\makeatother

\addcontentsline{toc}{section}{Conclusion}
\section{Conclusion}
\label{section: Conclusion}

A frequency domain analysis and a decomposition of all in-loop coordinates associated to {\it TM1} has been presented in order to highlight the {\it DFACS} performance of the {\it LISA Pathfinder} mission. The stability of the {\it LISA Pathfinder} platform, with respect to a local geodesic, has also been estimated.

A number of points can be concluded from this study:
\begin{itemize}

\item{The study has shown that the {\it LISA Pathfinder} platform has remarkable performance in terms of stability over all degrees of freedom. The privileged $X$ axis has outstanding performance but the other degrees of freedom show adequate performance which demonstrate the interest of such a platform for other applications. Improvements in some of the sensors and actuators could enhance this performance.}

\item{This study shows that the stability of {\it LISA Pathfinder}, in term of acceleration w.r.t local inertial reference frame, is sensitive to the \textit{GRS} sensing noise around $1 \si{\milli\hertz}$ and to {\it TMs} force noise at lower frequency. Above $0.1 \si{\hertz}$, the stability performances are impacted by the (micro-thruster) force noise and by the {\it DFACS} control loop}.

\item{Below $1 \si{\milli\hertz}$, the noise of the {\it Star Tracker} strongly impacts the performance of the system on all degrees of freedom. It should be noted however that, for {\it LISA}, several orders of magnitude improvements on attitude control performances are expected, benefiting from $~10^{-8} \si{\radian\persqrthz}$ precision attitude sensing with {\it DWS} on the incoming long-range laser beam \cite{amaro-seoane_laser_2017}, rather than the ($~10^{-4} \si{\radian\persqrthz}$) level achieved by the {\it LPF} star tracker at low frequency, around $0.1 \si{\milli\hertz}$.}

%Internal (capacitive) actuation noise levels do not appear to impact much on these  performances, except possibly for  $\eta$, $\phi$ and only in specific frequency ranges. Even in these ranges, there could be some ambiguity related to the  frequency dependence of the sensing noises. Because of this, only upper limits of the capacitive actuation noises on $\eta$, $\phi$ can be given by this study. We refer to \cite{Luigi_2017} and  \cite{Bill_2018} (\hl{reference for Bill's  papers }) for a more precise in-flight measurement of these noises.}

\item{The {\it LPF} {\it SSM} \cite{weyrich_modelling_2008} developed by the collaboration provides a  reliable description of the closed loop dynamics, showing  that the {\it LISA Pathfinder} system can be approximated by a linear system for frequencies lower than $0.2 Hz$}. Hence, the {\it State Space Model} has been used to estimate the stability of the {\it LISA Pathfinder} platform over a wide frequency range, highlighting its remarkable performances.

\item{The demonstrated reliability of the model is an item of interest for the upcoming task of extrapolating {\it LISA Pathfinder} results towards {\it LISA} simulations and design. Such work is ongoing and will be published in the near future.}

\item{The quality of the performances obtained by the {\it LISA Pathfinder} platform, with respect to the local geodesic, should therefore allow definition of similar platforms for other type of space-based measurements}.

\end{itemize}

\makeatletter

% Search and replace example with OgreKit:
% Search: \\vec{(?<i>\w)}_\\indice{(?<o>\w/\w)}
% Replace: \\myvec{\g<i>}{\g<o>}

\makeatother

\section{\label{S:acc} Acknowledgement}

This work has been made possible by the LISA Pathfinder mission, which is part of the
space-science programme of the European Space Agency.

The French contribution has been supported by the CNES (Accord Specific de projet
CNES 1316634/CNRS 103747), the CNRS, the Observatoire de Paris and the University
Paris-Diderot. E.~Plagnol and H.~Inchausp\'{e} would also like to acknowledge the
financial support of the UnivEarthS Labex program at Sorbonne Paris Cit\'{e}
(ANR-10-LABX-0023 and ANR-11-IDEX-0005-02).

The Albert-Einstein-Institut acknowledges the support of the German Space Agency,
DLR. The work is supported by the Federal Ministry for Economic Affairs and Energy
based on a resolution of the German Bundestag (FKZ 50OQ0501 and FKZ 50OQ1601). 

The Italian contribution has been supported  by Agenzia Spaziale Italiana and Istituto
Nazionale di Fisica Nucleare.

The Spanish contribution has been supported by contracts AYA2010-15709 (MICINN),
ESP2013-47637-P, and ESP2015-67234-P (MINECO). M.~Nofrarias acknowledges support from
Fundacion General CSIC (Programa ComFuturo). F.~Rivas acknowledges an FPI contract
(MINECO).

The Swiss contribution acknowledges the support of the Swiss Space Office (SSO)
via the PRODEX Programme of ESA. L.~Ferraioli is supported by the Swiss National
Science Foundation.

The UK groups wish to acknowledge support from the United Kingdom Space Agency
(UKSA), the University of Glasgow, the University of Birmingham, Imperial College,
and the Scottish Universities Physics Alliance (SUPA).

J.\,I.~Thorpe and J.~Slutsky acknowledge the support of the US National Aeronautics
and Space Administration (NASA).
\appendix
\bibliographystyle{apsrev4-1}
%\bibliography{Bibliographie, Papier_LPF_Performances_DFACSNotes.bib}
\bibliography{Bibliographie}
%\printbibliography
\end{document}